\documentclass[twocolumn,final,times,sort&compress]{elsarticle}
\usepackage{graphicx,amssymb,amsmath}
\usepackage{caption}
\usepackage{tikz}
\usepackage{natbib}
\journal{Physica B}

\begin{document}

\begin{frontmatter}

\title{The influence of further-neighbor spin-spin interaction on a ground state  of 2D coupled spin-electron model in a magnetic field\tnoteref{grant}}
\tnotetext[grant]{This work was supported by the Slovak Research and Development Agency (APVV) under Grants No. APVV-16-0186, APVV-0097-12 and APVV-14-0878. The financial support provided by the VEGA under Grants No. 1/0043/16 and 2/0130/15 is also gratefully acknowledged. }
\author[UEFSAV]{Hana \v Cen\v carikov\'a\corref{coraut}} 
\cortext[coraut]{Corresponding author}
\ead{hcencar@saske.sk}
\author[UPJS]{Jozef Stre\v{c}ka}
\author[FUSAV]{Andrej Gendiar}
\author[UEFSAV]{Nat\'alia Toma\v sovi\v cov\'a}
\address[UEFSAV]{Institute  of  Experimental  Physics,  Slovak   Academy   of Sciences, Watsonova 47, 040 01 Ko\v {s}ice, Slovakia}
\address[UPJS]{Institute of Physics, Faculty of Science, P. J. \v{S}af\'{a}rik University, Park Angelinum 9, 04001 Ko\v{s}ice, Slovakia}
\address[FUSAV]{Institute of Physics, Slovak Academy of Sciences, D\'ubravsk\'a cesta 9, SK-845 11, Bratislava, Slovakia}

\begin{abstract}
An exhaustive ground-state analysis of extended two-dimensional (2D) correlated spin-electron model consisting of the Ising spins localized on nodal lattice sites and mobile electrons delocalized over pairs of decorating sites is performed within the framework of rigorous analytical calculations. The investigated model, defined on an arbitrary 2D doubly decorated lattice, takes into account the kinetic energy of mobile electrons, the nearest-neighbor Ising coupling between the localized spins and mobile electrons, the further-neighbor Ising coupling between the localized spins and the Zeeman energy. The ground-state phase diagrams are examined for a wide range of model parameters for both ferromagnetic as well as antiferromagnetic interaction between the nodal Ising spins and non-zero value of external magnetic field. It is found that non-zero values of further-neighbor interaction leads to a formation of new quantum states as a consequence of competition between all considered interaction terms. Moreover, the new quantum states are accompanied with different magnetic features and thus, several kinds of discontinuous field-driven  phase transitions are observed.
\end{abstract}


\end{frontmatter}

\section{Introduction}

During several last decades a considerable amount of effort has been devoted to the investigation of coupled spin-electron systems due to the fact, that such materials exhibit a wide range of unconventional properties~\cite{Wachter,Cheong} with a direct application in the real life. Their application potential makes such materials very attractive for physicists as well as engineers,  but in spite of their enormous effort, the exhaustive understanding of driven mechanisms in such complex systems has not been achieved so far. In general, it is assumed that the origin of mentioned collective phenomena 
arises from a competition between electron motion and magnetic behavior~\cite{Velu,Baibich}, however, the importance of  selected contributions is still highly debated. From the theoretical point of view, the special interest has been devoted to the relevance of additional interaction terms, which are often neglected in the first approach analysis but could be responsible for the new interesting behavior as exemplified in Refs.~\cite{Motrunich,Cenci3,Liu}. 
\newline\hspace*{0.2cm}
In the present paper we investigate the role of direct spin-spin interaction on the formation of magnetic order, where we suppose that its presence fundamentally contributes to a magnetic diversity of real materials~\cite{Carlin}. As known, the diversity  of magnetic states is highly desired in various field sensing devices and/or should be the base for presence of huge magnetocaloric effect significant for the refrigeration purposes. Consequently, their detailed examination is,  therefore, very valuable.
For the theoretical analysis we propose a relatively simple extended spin-electron model on an arbitrary doubly decorated lattices with the localized Ising spins and delocalized mobile electrons, the simplified versions of which have been previously studied in 1D~\cite{Cisarova1,Cisarova2} as well as 2D cases~\cite{Doria,Cenci1,Cenci2}. In spite of the model simplicity, the previous results point to the model convenience and present a good agreement with experimental observations. 
Our further analysis is primarily focused on the examination of the magnetic ground-state phase diagrams under the influence of external magnetic field, where an exhaustive description of stable magnetic states is precisely done. Besides, we accurately examine the stability area of each phase and define the exact boundary conditions among them. Finally, we detect the presence of field-driven discontinuous phase transition and specify the conditions of their existence.
\newline\hspace*{0.2cm}
The paper is organized as follows. In Section~\ref{s2} we briefly describe the investigated model and derive the eigenvalues of bond Hamiltonian as a necessary step in  determination of a ground-state energy. The most interesting results with the corresponding discussion are presented in Section~\ref{s3} and last, a few conclusions together with future outlooks are collected in Section~\ref{s4}.

\section{Model and Method}
\label{s2}
The proposed  coupled spin-electron model on doubly decorated planar lattice is formed by immobile Ising spins localized at each nodal lattice site and by mobile electrons delocalized over the pairs of sites decorating each bond. The energy terms occurring in the model Hamiltonian correspond to the kinetic energy of mobile electrons, the Ising interaction  between the mobile electrons and their nearest-neighbor Ising spins, as well as, the Ising interaction between the nearest-neighbor Ising spins. Of course, the Zeeman energy term must be included to study the effect of external magnetic field. The mutual commutativity between different bond Hamiltonians  $\hat{\cal H}_k$ enable us to rewrite the total Hamiltonian  $\hat{\cal H}$ to the more convenient form: $\hat{\cal H}=\sum_{k=1}^{Nq/2} \hat{\cal H}_k$, where $N$ is a total number of all nodal sites and $q$ is the coordination number. Then the bond Hamiltonian can be defined as:
\begin{eqnarray}
\hat{\cal H}_k=\!\!\!&-&\!\!\!t(\hat{c}^\dagger_{k_1,\uparrow}\hat{c}_{k_2,\uparrow}+\hat{c}^\dagger_{k_1,\downarrow}\hat{c}_{k_2,\downarrow}+
\hat{c}^\dagger_{k_2,\uparrow}\hat{c}_{k_1,\uparrow}+\hat{c}^\dagger_{k_2,\downarrow}\hat{c}_{k_1,\downarrow})
\nonumber
\\
\!\!\!&-&\!\!\!J\hat\sigma^z_{k_1}(\hat{n}_{k_1,\uparrow}-\hat{n}_{k_1,\downarrow})-
J\hat\sigma^z_{k_2}(\hat{n}_{k_2,\uparrow}-\hat{n}_{k_2,\downarrow})
\nonumber\\
\!\!\!&-&\!\!\!h(\hat{n}_{k_1,\uparrow}-\hat{n}_{k_1,\downarrow})-
h(\hat{n}_{k_2,\uparrow}-\hat{n}_{k_2,\downarrow})
\label{eq1}\\
\!\!\!&-&\!\!\!\frac{h}{q}(\hat\sigma^z_{k_1}+\hat\sigma^z_{k_2})-J'\hat\sigma^z_{k_1}\hat\sigma^z_{k_2}-\mu(\hat{n}_{k_1}+\hat{n}_{k_2})\;,\nonumber
\end{eqnarray}
where the symbols $\hat{c}^\dagger_{k_{\alpha},\gamma}$/$\hat{c}_{k_{\alpha},\gamma}$ ($\alpha$=1,2; $\gamma=\uparrow,\downarrow$) denote the creation/annihilation fermionic operators of the mobile electron and $\hat{n}_{k_{\alpha},\gamma}=\hat{c}^\dagger_{k_{\alpha},\gamma}\hat{c}_{k_{\alpha},\gamma}$ as well as $\hat{n}_{k_{\alpha}}=\sum_{\{\gamma\}}\hat{n}_{k_{\alpha},\gamma}$ are the corresponding number operators.  $\hat\sigma^z_{k_{\alpha}}$ denotes the $z$-component of the Pauli operator with the eigenvalues $\sigma=\pm1$. The first term in Eq.~(\ref{eq1}) corresponds to the kinetic energy of mobile electrons delocalized over a couple of decorating  sites $k_1$ and $k_2$ from the $k$-th dimer with the hopping amplitude $t$. The second  and the third terms represent the Ising interaction between the mobile electrons and their nearest-neighbors Ising spins described by the parameter $J$. The next three terms in the Eq.~(\ref{eq1}) correspond to the energy contribution induced by the external magnetic field acting on the localized  as well as delocalized particles and the term $J'$ denotes the Ising interaction between the nearest-neighbor Ising spins. Finally, $\mu$ is a chemical potential of the mobile electrons. 
                                                     
To perform an exhaustive analysis of the ground state it is necessary to obtain the eigenvalues of the bond Hamiltonian. The bond Hamiltonian $\hat{\cal H}_k$ can be divided  into several disjoint blocks ${\cal H}_k(n_k)$ due to the commutativity of $\hat{\cal H}_k$ with the number operator of mobile electrons per bond $\hat{n}_k$ and the calculation  procedure is significantly simplified. Subsequently, the sixteen different eigenvalues $E_k$, corresponding to the different electron fillings have been obtained:
\begin{eqnarray}
\begin{array}{ll}
n_k=0:& E_{k_1}=R,\\
n_k=1:& E_{k_2,\,k_3}= \pm JP+Q/2\pm h+R-\mu,\\
&E_{k_4,\,k_5}= \pm JP-Q/2\pm h+R-\mu,\\
n_k=2: & E_{k_6,\,k_7}=\pm 2JP\pm 2h+R-2\mu,\\
 & E_{k_8}=E_{k_9}=R-2\mu,\\
 & E_{k_{10},\,k_{11}}=\pm Q+R-2\mu,\\
n_k=3: & E_{k_{12},\,k_{13}}= \pm JP+Q/2\pm h+R-3\mu,\\
& E_{k_{14},\,k_{15}}= \pm JP-Q/2\pm h+R-3\mu,\\
n_k=4: &E_{k_{16}}=R-4\mu,
\end{array}
\label{eq4}
\end{eqnarray}
where $P=(\sigma_{k_1}+\sigma_{k_2})/2$, $Q=\sqrt{J^2(\sigma_{k_1}-\sigma_{k_2})^2+4t^2}$ and $R=-J\sigma_{k_1}\sigma_{k_2}-hL/q$.

\section{Results and discussion}
\label{s3}
In this section we present the most interesting results obtained from the ground-state analysis of the model (\ref{eq1}) in the presence of external magnetic field focusing on the diversity of stable magnetic structures.  First of all, it should be mentioned that the absolute value of the coupling constant $J$ between the localized spins and mobile electrons is set to unity and  all others parameters will be normalized with respect to this coupling.  In addition, the applied magnetic field is always chosen positive, i.e. $h>0$  and the coordination number $q=4$ is assumed.
\newline\hspace*{0.2cm} 
To investigate the ground-state energy of the model (\ref{eq1}) all 64  possible magnetic states derived from Eq.~(\ref{eq4}) by considering four available combinations of two Ising spins must be taken into account.  Fortunately, out of the whole investigated ensemble, only 15 different phases may become ground state. These phases together with their energies are collected in Tab.~{\ref{tab1}}.  
\begin{table*}[t!]
\begin{center}
\resizebox{1\textwidth}{!} {
\begin{tabular}{l||l|l}
Electron filling & Eigenvalue (${E}$)& Eigenvector \\
\hline\hline
 $\begin{array}{c}
    \rho=0
   \end{array}$       
&
$\displaystyle
\begin{array}{l}
    {E}(\mbox{0}_1)=-2h/q-J'\\
    {E}(\mbox{0}_2)=J'
 \end{array}   $
&
$\displaystyle
\begin{array}{l}
   |\mbox{0}_1\rangle=\prod_{k=1}^{Nq/2}|1\rangle_{\sigma_{k_1}}\otimes |0,0\rangle_{k}\otimes|1\rangle_{\sigma_{k_2}} \\
        |\mbox{0}_2\rangle=\prod_{k=1}^{Nq/2}|1\rangle_{\sigma_{k_1}}\otimes |0,0\rangle_{k}\otimes|-1\rangle_{\sigma_{k_2}}
\end{array}   $
\\
\hline
 $\begin{array}{c}
    \rho=1
   \end{array}$       
&
$\displaystyle
\begin{array}{l}
    {E}(\mbox{I}_1)=-J-J'-h-2h/q-t-\mu\\
    {E}(\mbox{I}_2)=J-J'-h+2h/q-t-\mu\\
    {E}(\mbox{I}_3)=J'-h-\sqrt{J^2+t^2}-\mu
   \end{array}   $
&
$\displaystyle
\begin{array}{l}
   |\mbox{I}_1\rangle=\prod_{k=1}^{Nq/2}|1\rangle_{\sigma_{k_1}}\otimes \frac{1}{\sqrt{2}}\left(
|\!\uparrow,0\rangle_k+|0,\uparrow\rangle_k\right)\otimes|1\rangle_{\sigma_{k_2}}\\
   |\mbox{I$_2$}\rangle=\prod_{k=1}^{Nq/2}|-1\rangle_{\sigma_{k_1}}\otimes \frac{1}{\sqrt{2}}\left(
|\!\uparrow,0\rangle_k+|0,\uparrow\rangle_k\right)\otimes|-1\rangle_{\sigma_{k_2}}\\
     |\mbox{I$_3$}\rangle=\prod_{k=1}^{Nq/2}|1\rangle_{\sigma_{k_1}}\otimes \left(\alpha
|\!\uparrow,0\rangle_k+\beta|0,\uparrow\rangle_k\right)\otimes|-1\rangle_{\sigma_{k_2}}
  \end{array}   $
 \\
\hline
 $\begin{array}{c}
    \rho=2
   \end{array}$       
&
$\displaystyle
\begin{array}{l}
    {E}(\mbox{II}_1)=-2J-J'-2h-2h/q-2\mu\\
    {E}(\mbox{II}_2)=-J'-2h/q-2t-2\mu\\
    {E}(\mbox{II}_3)=2J-J'-2h+2h/q-2\mu\\
    {E}(\mbox{II}_4)=J'-2h-2\mu\\
    {E}(\mbox{II}_5)=J'-2\sqrt{J^2+t^2}-2\mu
\end{array}   $
&
$\displaystyle
\begin{array}{l}
    |\mbox{II$_1$}\rangle=\prod_{k=1}^{Nq/2}|1\rangle_{\sigma_{k_1}}\otimes |\!\uparrow,\uparrow\rangle_k\otimes|1\rangle_{\sigma_{k_2}}\\
|\mbox{II}_2\rangle=\prod_{k=1}^{Nq/2}|1\rangle_{\sigma_{k_1}}\otimes\frac{1}{2}\left[|\!\uparrow,\downarrow\rangle_k-|\!\downarrow,\uparrow\rangle_k
+|\!\uparrow\downarrow,0\rangle_k+|0,\uparrow\downarrow\rangle_k\right]\otimes|1\rangle_{\sigma_{k_2}}    \\
 |\mbox{II$_3$}\rangle=\prod_{k=1}^{Nq/2}|-1\rangle_{\sigma_{k_1}}\otimes |\!\uparrow,\uparrow\rangle_k\otimes|-1\rangle_{\sigma_{k_2}}\\
  |\mbox{II$_4$}\rangle=\prod_{k=1}^{Nq/2}|1\rangle_{\sigma_{k_1}}\otimes |\!\uparrow,\uparrow\rangle_k\otimes|-1\rangle_{\sigma_{k_2}}\\
 |\mbox{II$_5$}\rangle=\prod_{k=1}^{Nq/2}|1\rangle_{\sigma_{k_1}}\otimes \left[a|\!\uparrow,\downarrow\rangle_k+b|\!\downarrow,\uparrow\rangle_k
+c\left(|\!\uparrow\downarrow,0\rangle_k+|0,\uparrow\downarrow\rangle_k\right)\right]\otimes|-1\rangle_{\sigma_{k_2}}  
   \end{array}  $
\\
\hline
 $\begin{array}{c}
    \rho=3
   \end{array}$       
&
$\displaystyle
\begin{array}{l}
 {E}(\mbox{III}_1)=-J-J'-h-2h/q-t-3\mu\\
    {E}(\mbox{III}_2)=J-J'-h+2h/q-t-3\mu\\
    {E}(\mbox{III}_3)=J'-h-\sqrt{J^2+t^2}-3\mu
  \end{array}   $
&
$\displaystyle
\begin{array}{l}
   |\mbox{III$_1$}\rangle=\prod_{k=1}^{Nq/2}|1\rangle_{\sigma_{k_1}}\otimes \frac{1}{\sqrt{2}}\left(|\!\uparrow\downarrow,\uparrow\rangle_k-|\!\uparrow,\uparrow\downarrow\rangle_k\right)\otimes|1\rangle_{\sigma_{k_2}}\\
     |\mbox{III$_2$}\rangle=\prod_{k=1}^{Nq/2}|-1\rangle_{\sigma_{k_1}}\otimes \frac{1}{\sqrt{2}}\left(|\!\uparrow\downarrow,\uparrow\rangle_k-|\!\uparrow,\uparrow\downarrow\rangle_k\right)\otimes|-1\rangle_{\sigma_{k_2}}\\
     |\mbox{III$_3$}\rangle=\prod_{k=1}^{Nq/2}|1\rangle_{\sigma_{k_1}}\otimes\left(\beta|\!\uparrow\downarrow,\uparrow\rangle_k-\alpha|\!\uparrow,\uparrow\downarrow\rangle_k\right)\otimes|-1\rangle_{\sigma_{k_2}}
  \end{array}   $
 \\
\hline
 $\begin{array}{c}
    \rho=4
   \end{array}$       
&
$\displaystyle
\begin{array}{l}
    {E}(\mbox{IV}_1)=-2h/q-J'-4\mu\\
      {E}(\mbox{IV}_2)=J'-4\mu
 \end{array}   $
&
$\displaystyle
\begin{array}{l}
   |\mbox{IV}_1\rangle=\prod_{k=1}^{Nq/2}|1\rangle_{\sigma_{k_1}}\otimes |\!\uparrow\downarrow,\uparrow\downarrow\rangle_k\otimes|1\rangle_{\sigma_{k_2}}\\
        |\mbox{IV}_2\rangle=\prod_{k=1}^{Nq/2}|1\rangle_{\sigma_{k_1}}\otimes |\!\uparrow\downarrow,\uparrow\downarrow\rangle_k\otimes|-1\rangle_{\sigma_{k_2}}
\end{array}   $
 \\
\hline
\end{tabular}}
\end{center}
\caption{The list of eigenvalues and eigenvectors forming individual ground states. The probability amplitudes $\alpha$ and $\beta$ used in the notation of the eigenvectors  $|$I$_3\rangle$ / $|$III$_3\rangle$ have the explicit forms: $\alpha=\frac{\left(\sqrt{J^2+t^2}+J\right)}{\sqrt{2\left(J^2+t^2+J\sqrt{J^2+t^2}\right)}}$ and $\beta=\frac{t}{\sqrt{2\left(J^2+t^2+J\sqrt{J^2+t^2}\right)}}$, while probability amplitudes $a$, $b$, and $c$ used in the notation of the eigenvector  $|$II$_5\rangle$ have the explicit forms: $a=\frac{J+\sqrt{J^2+t^2}}{2\sqrt{J^2+t^2}}$, $b=\frac{-(\sqrt{J^2+t^2}-J)}{2\sqrt{J^2+t^2}}$, and $c=\frac{t}{2\sqrt{J^2+t^2}}$.}
\label{tab1}
\end{table*}   
As one can see, the model can stabilize both the ferromagnetic (F) as well as antiferromagnetic (AF) type of long-range ordering in both subsystems for an arbitrary integer electron concentration. In comparison with the previous studies of identical model~\cite{Cenci2,Cenci4}, the mutual influence of all present interactions results into existence of novel magnetic phases, which are absent in the model without the magnetic field or the further-neighbor interaction $J'$. In this context, there arises a question whether all 15 ground states could be achieved by a simple modulation of just one external parameter, for instance, the magnetic field $h$. If there existed a conformable answer, then there would  exist relatively simple way how to alter various magnetic states with a direct utilization in the real life.

Let us analyze the obtained results in detail,  dividing them according to the type of the spin-electron interaction $J$ (the F type if $J>0$ and the AF type otherwise) for both, the F as well as AF type of the further-neighbor interaction $J'$.

\subsection{The ferromagnetic case $J>0$}
\begin{figure*}[t!]
\begin{center}
\includegraphics[width=0.3\textwidth,trim=0 0 1.3cm 0.5cm, clip]{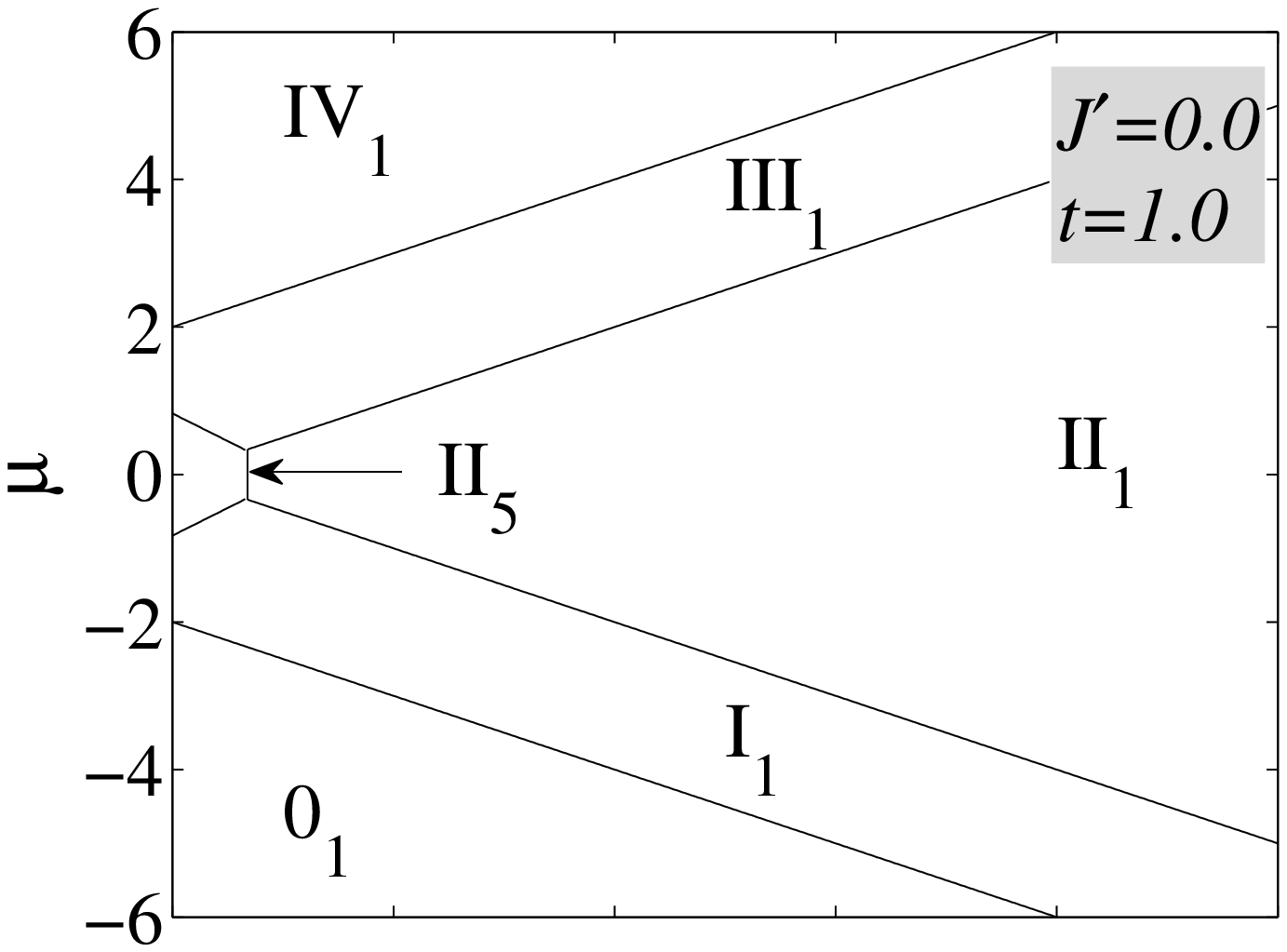}
\includegraphics[width=0.3\textwidth,trim=0 0 1.3cm 0.5cm, clip]{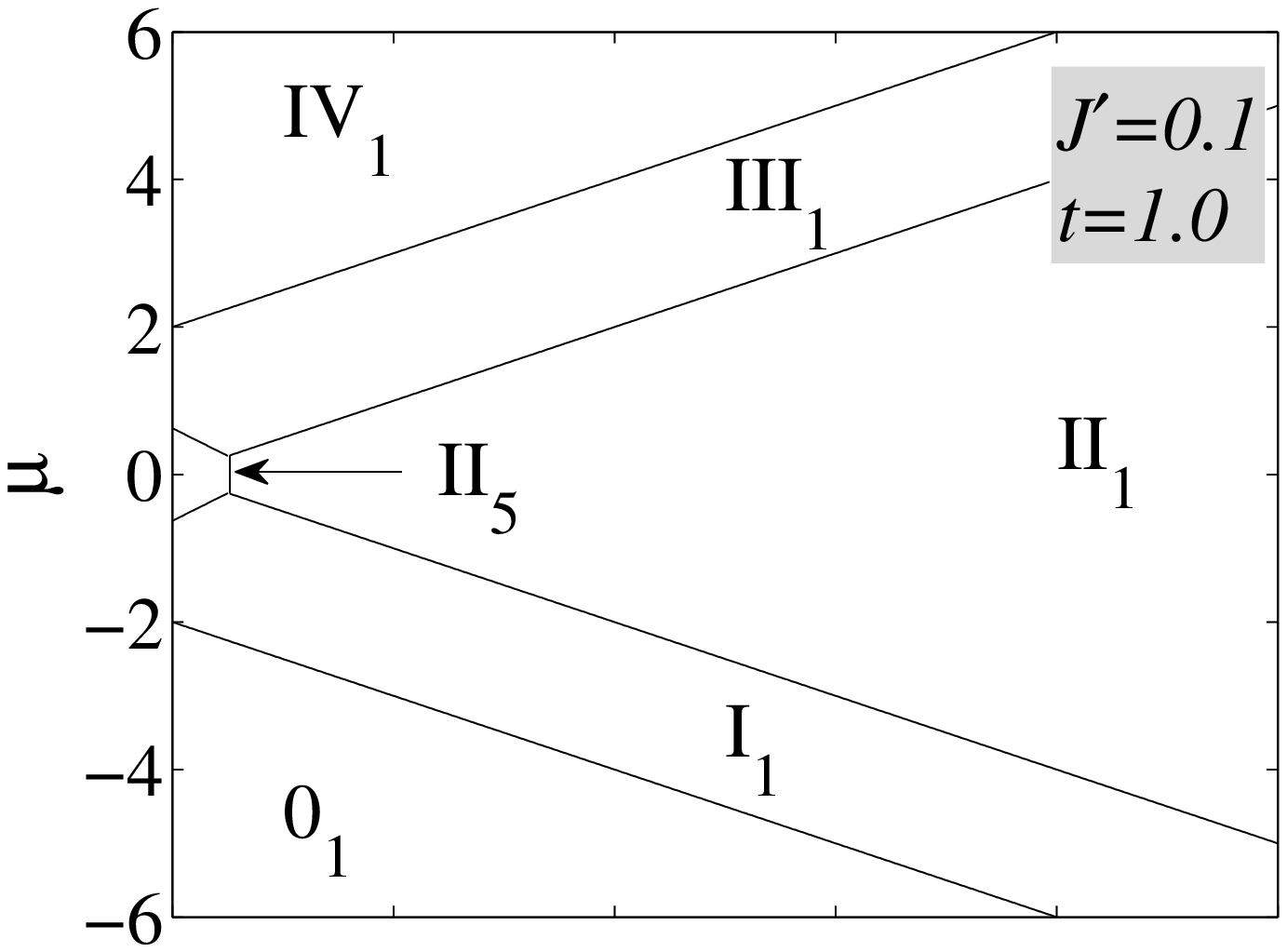}
\includegraphics[width=0.3\textwidth,trim=0 0 1.3cm 0.5cm, clip]{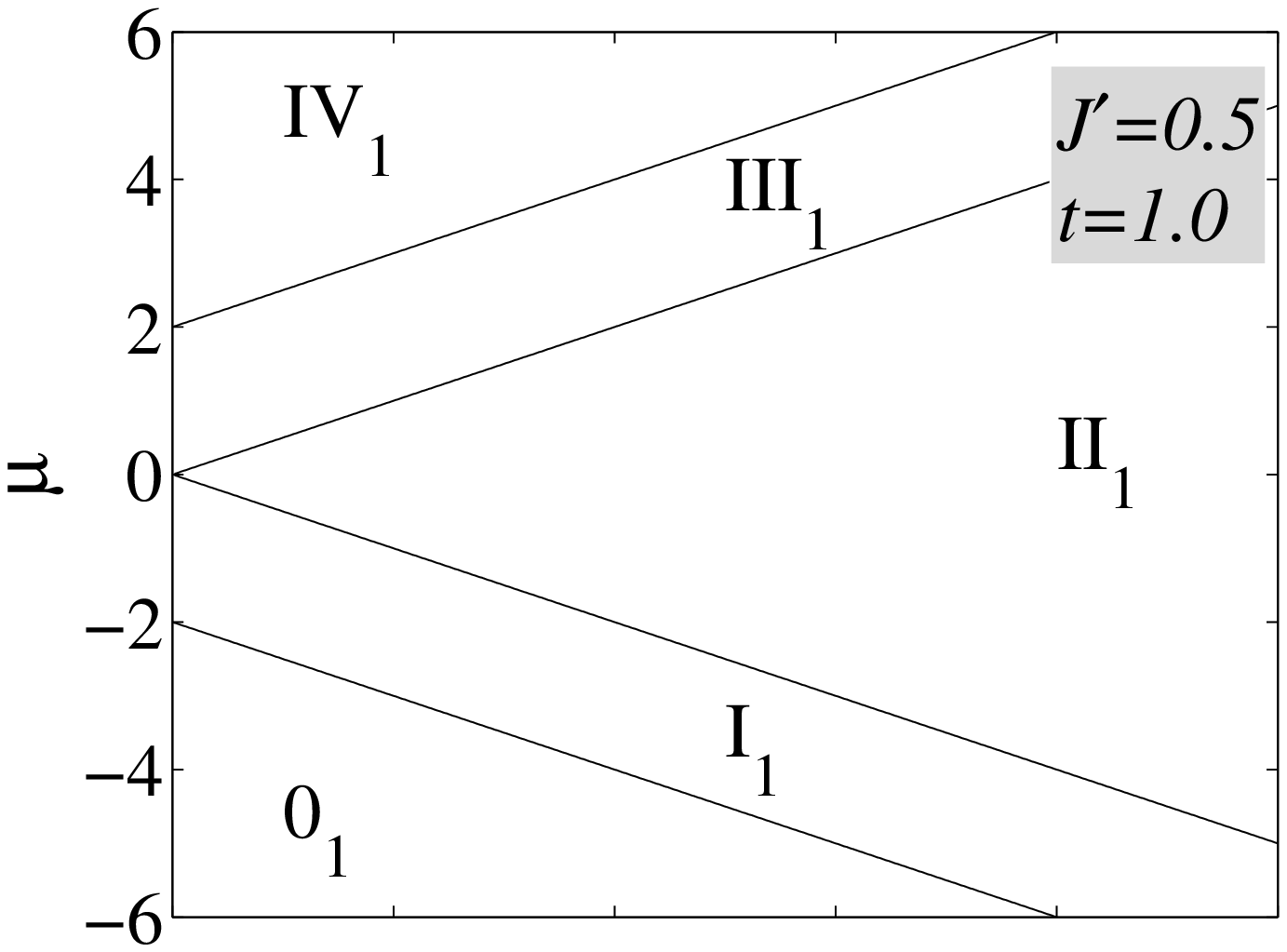}
\\\vspace*{-0.39cm}
{\includegraphics[width=0.3\textwidth,trim=0 0 1.3cm 0.5cm, clip]{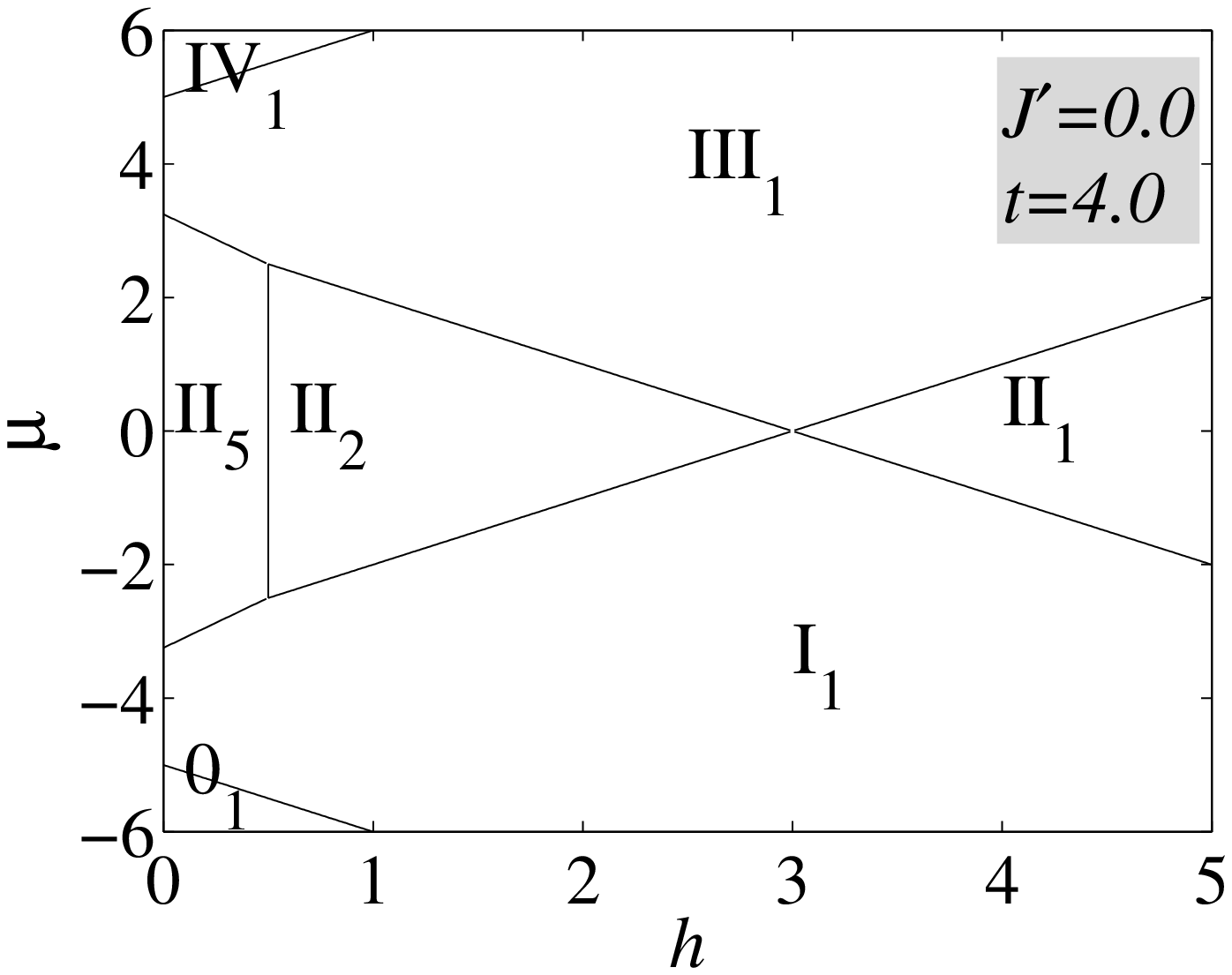}}
\includegraphics[width=0.3\textwidth,trim=0 0 1.3cm 0.5cm, clip]{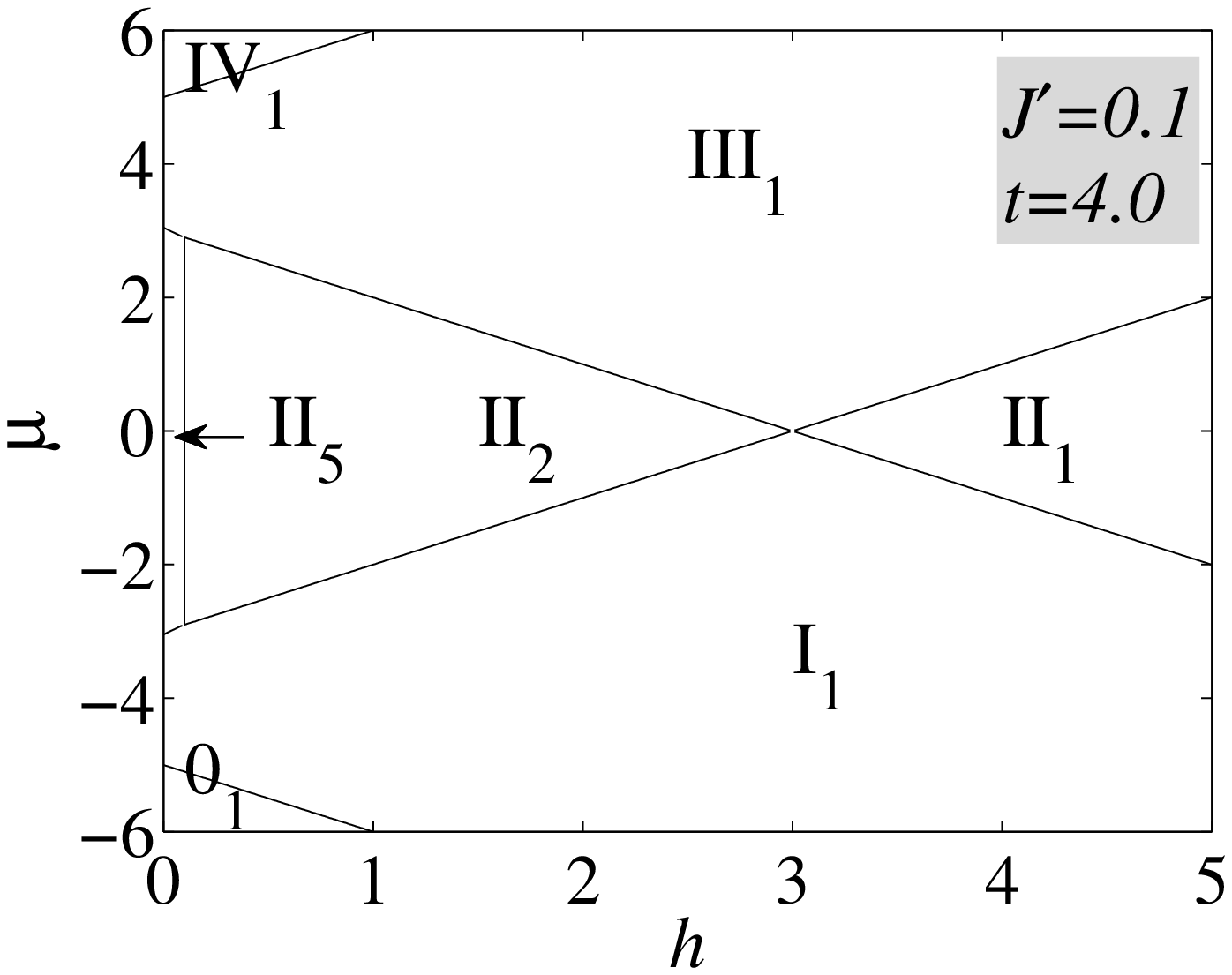}
\includegraphics[width=0.3\textwidth,trim=0 0 1.3cm 0.5cm, clip]{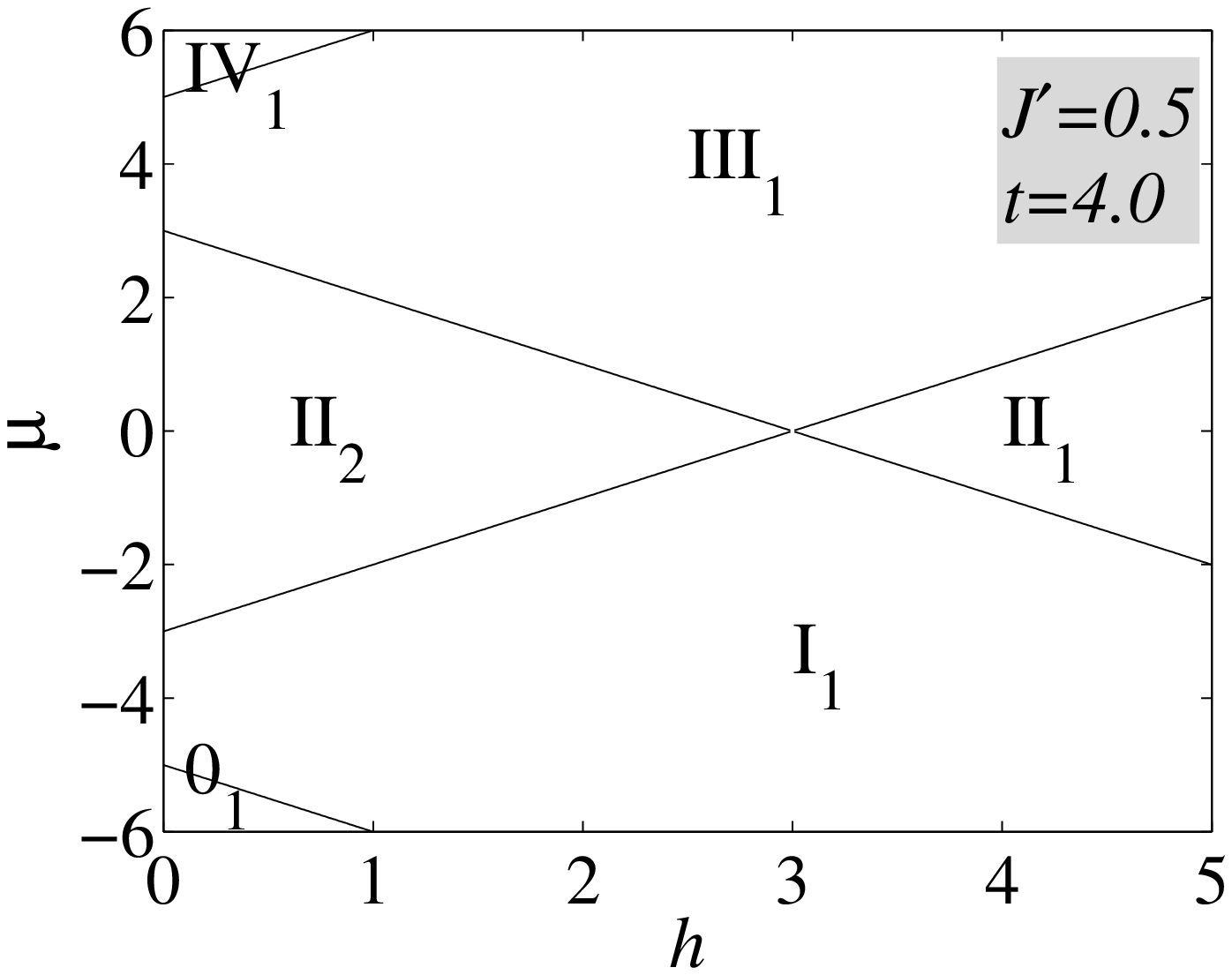}
\caption{\small Ground-state phase diagrams in the $\mu$-$h$ plane for $J=1$ and selected values of $J'\geq 0$ and $t$.}
\label{fig2}
\end{center}
\end{figure*}

As observed previously for the special case $J'$~=~0~\cite{Cenci4}, the F interaction $J>0$ in combination with the electron hopping $t$ and external magnetic field $h$  results in three different types of magnetic phase diagrams depending on a relative strength of the hopping term. 
An arbitrary non-zero field favors the F spin-electron state in an uncompensated and empty/fully electron occupancy, while in the half-filled band case the magnetic field  enforces discontinuous phase transitions. Of course, only the spin subsystem is F in the case $n_k=0$ since the electron subsystem is empty, similarly as in the fully occupied case $n_k=4$ when the electron subsystem resides in a non-magnetic ionic state. The F interaction $J'>0$ has only an insignificant influence on the phase stability, however, it markedly influences presence/existence of the field-driven phase transitions detected  in the half filling. This fact is clearly visible in Fig.~\ref{fig2}, 
where the respective phase boundaries are given by the following expressions:
\hspace*{-0.2cm}
\begin{eqnarray}
\begin{array}{ll}
{\rm 0}_1{\rm -I}_1\;/\;{\rm III}_1{\rm -IV_1}:&\!\!\!\!\mu\!=\!(-1)^{u}(-J-h-t),\\
{\rm I}_1{\rm -II}_1\;/\;{\rm II}_1{\rm -III}_1:&\!\!\!\! \mu\!=\!(-1)^{u}(-J-h+t),\\ 
{\rm I}_1{\rm -II}_2\;/\;{\rm II}_2{\rm -III}_1:&\!\!\!\! \mu\!=\!(-1)^{u}(J+h-t),\\
{\rm I}_1{\rm -II}_5\;/\;{\rm II}_5{\rm -III}_1:&\!\!\!\! \mu\!=\!(-1)^{u}\left(J+h+t\right.\\
&\left.+2(J'+h/q-\sqrt{J^2+t^2})\right). 
\end{array}\label{eqA1}
\hspace{-1cm}\end{eqnarray} 
It should be mentioned that $u=1$ if the $n_k$ of both adjacent phases are less or equal to two ($n_k\leq 2$) or $u=2$ otherwise.
As one can see, almost all borders are $J'$-invariant and hence only the phase boundaries I$_1$/III$_1$-II$_5$  depend on the spin-spin interaction $J'$. It can be found from Fig.~\ref{fig2} that the F coupling $J'>0$ dramatically reduces the value of applied magnetic field at which a field-induced phase transition occurres. The exception from this rule represents the phase boundary II$_1$-II$_2$, which is completely independent of $J'$ :
\begin{eqnarray}
\begin{array}{ll}
{\rm II}_1{\rm -II}_2:& h\!=\!-J+t,\\
{\rm II}_1{\rm -II}_5:& h/q\!=\!\left(-J-J'+\sqrt{J^2+t^2}\right)/(q+1),\label{eqA2}   \\
{\rm II}_2{\rm -II}_5:& h/q\!=\!-t-J'+\sqrt{J^2+t^2}.
\end{array}
\end{eqnarray}
The reduction of transition fields for the phase boundaries II$_1$-II$_3$ and II$_2$-II$_5$ relates with the fact that the F coupling $J'>0$ favors  parallel orientation of localized spins with respect to the antiparallel one. Depending on a relative strength of the hopping term, which prefers an opposite (antiparallel) orientation of electrons, either the state II$_1$ or state II$_2$ becomes dominant. To conclude, the F interaction $J'>0$ is not able to generate new magnetic phases, it only stabilizes/reduces the ones existing in the former phase diagrams.
\begin{figure*}[t!]
\begin{center}
\includegraphics[width=0.3\textwidth,trim=0 0 1.3cm 0.5cm, clip]{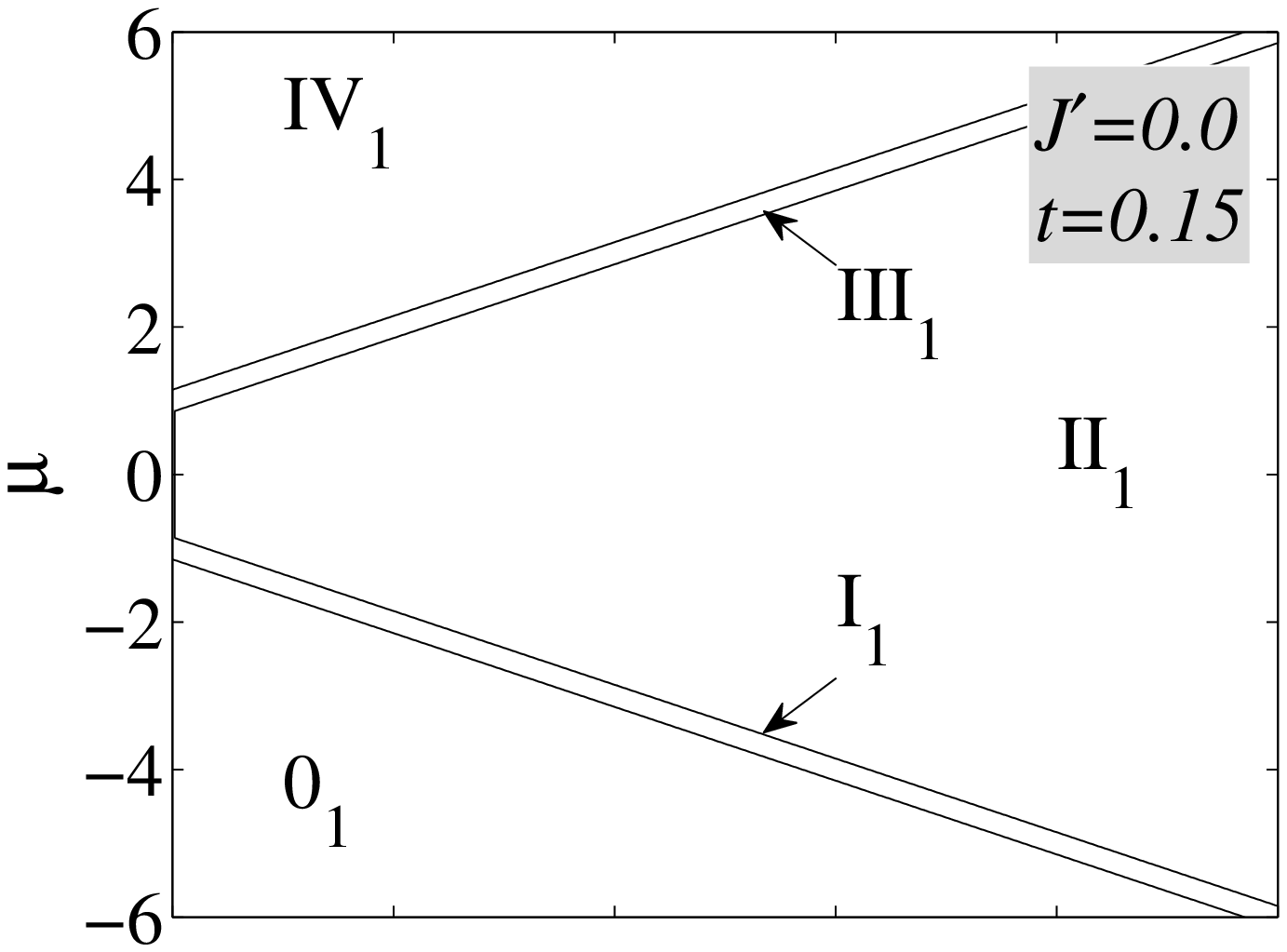}
\includegraphics[width=0.3\textwidth,trim=0 0 1.3cm 0.5cm, clip]{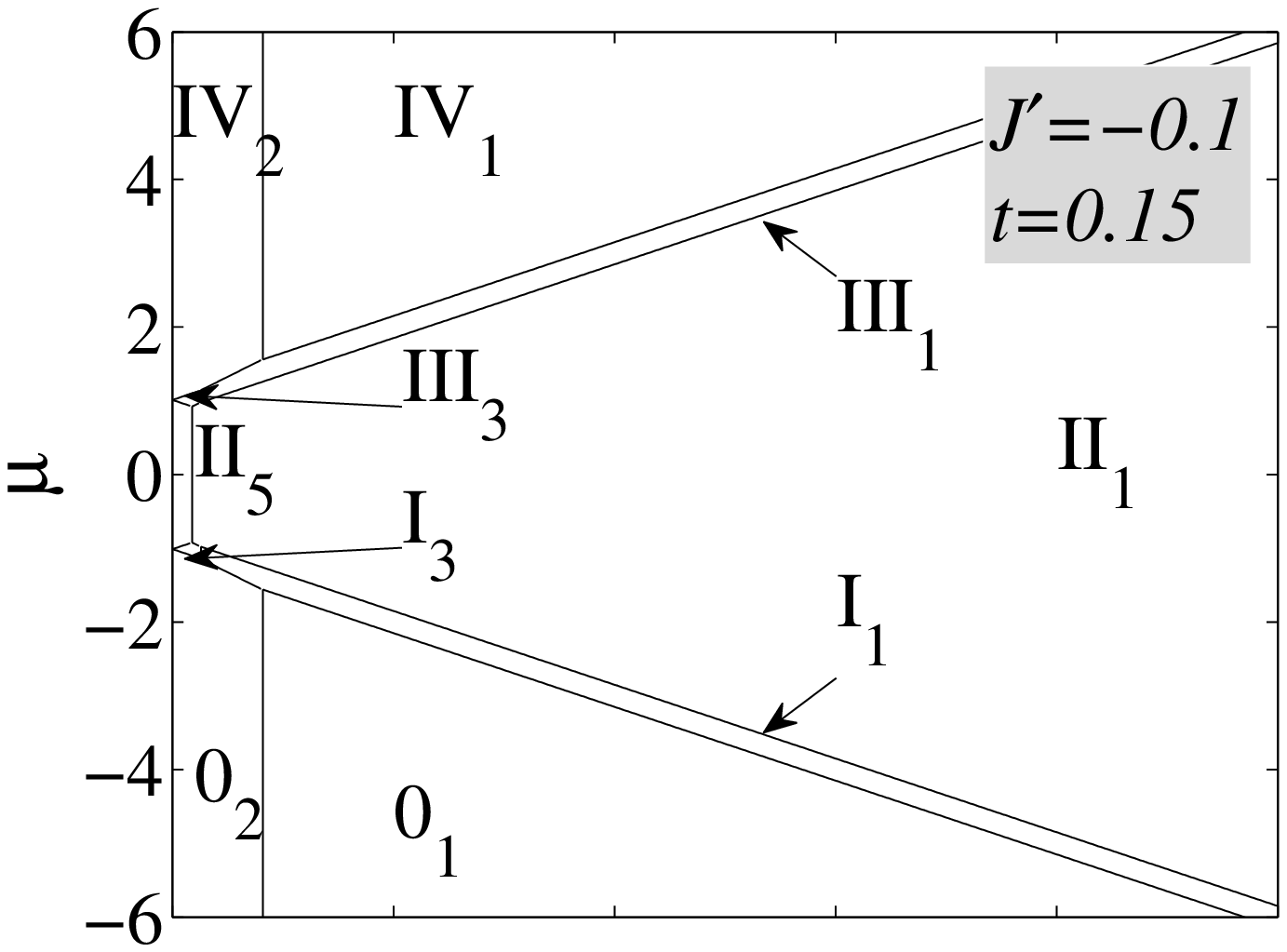}
\includegraphics[width=0.3\textwidth,trim=0 0 1.3cm 0.5cm, clip]{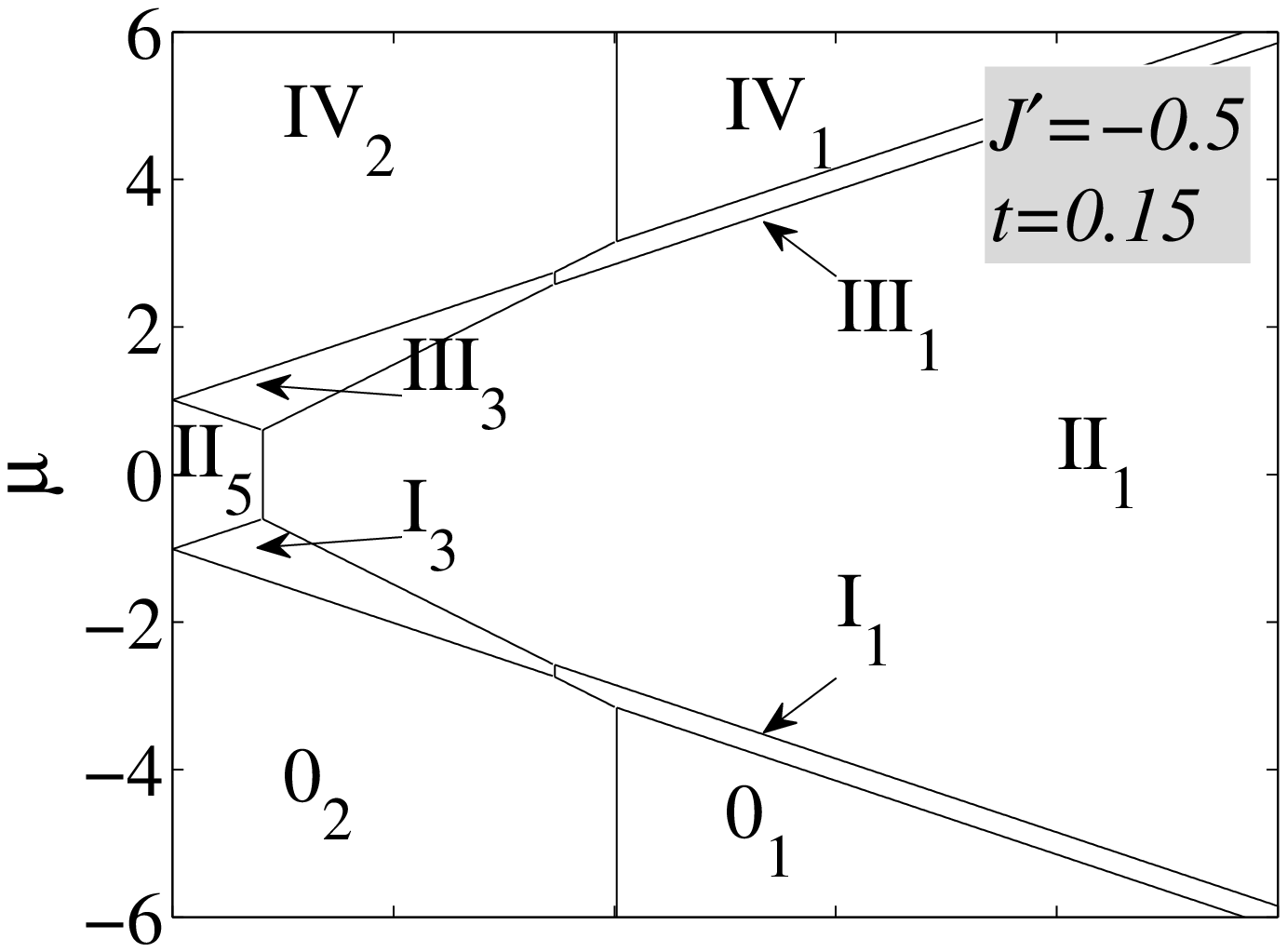}
\\\vspace*{-0.39cm}
\includegraphics[width=0.3\textwidth,trim=0 0 1.3cm 0.5cm, clip]{fsces_fdhjj0c0t1c0.eps}
\includegraphics[width=0.3\textwidth,trim=0 0 1.3cm 0.5cm, clip]{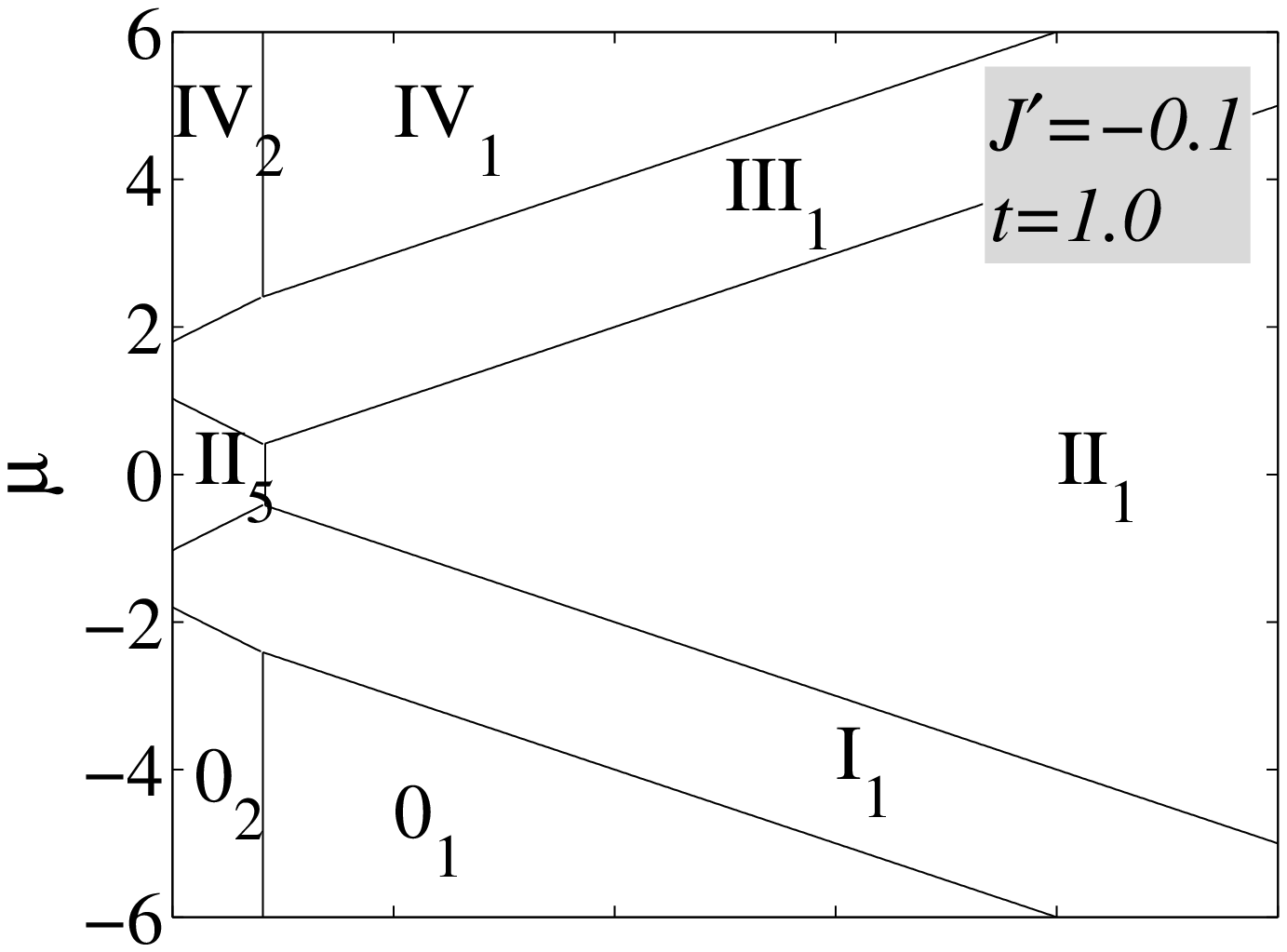}
\includegraphics[width=0.3\textwidth,trim=0 0 1.3cm 0.5cm, clip]{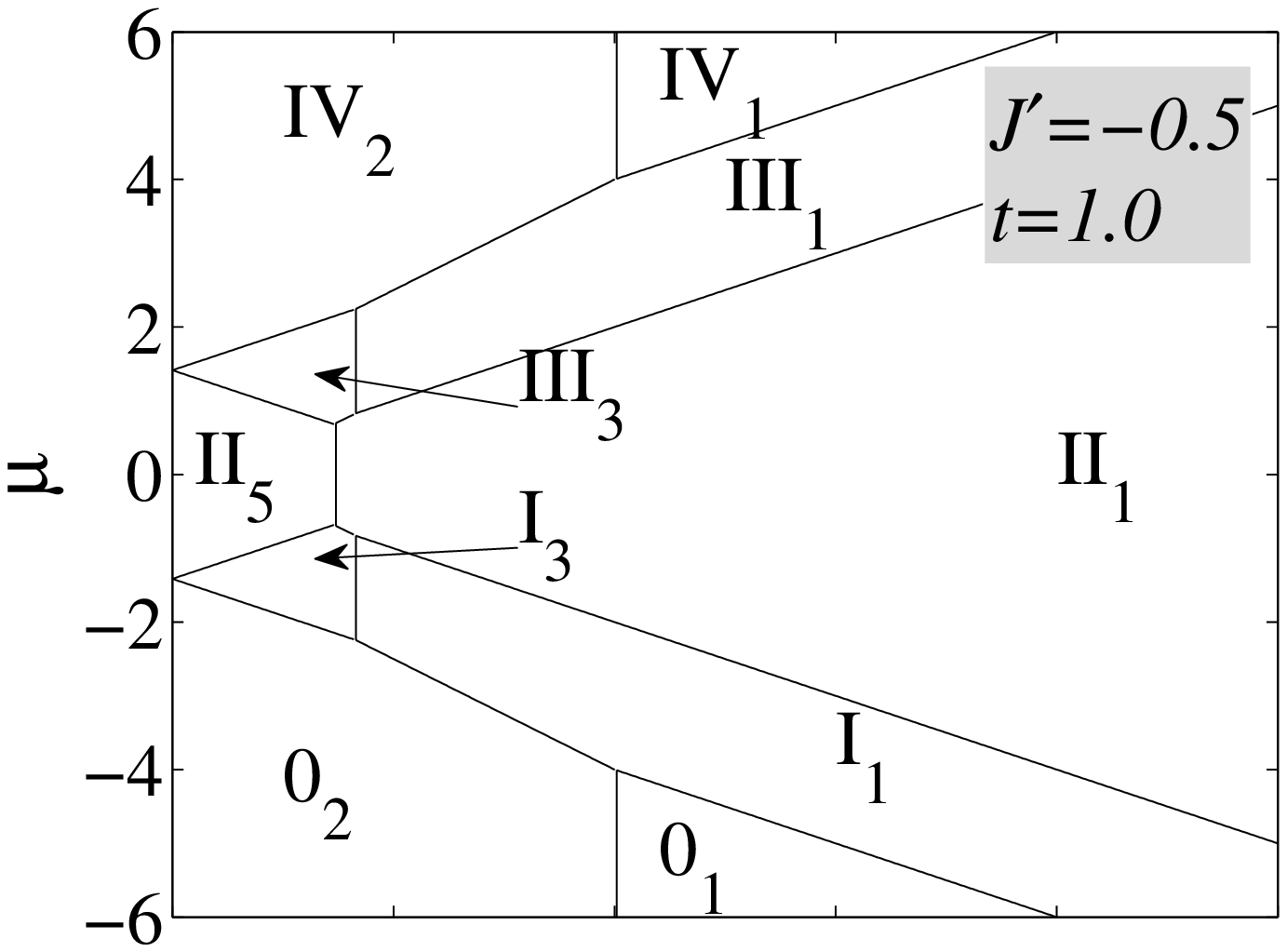}
\\\vspace*{-0.39cm}
\includegraphics[width=0.3\textwidth,trim=0 0 1.3cm 0.5cm, clip]{fsces_fdhjj0c0t4c0.eps}
\includegraphics[width=0.3\textwidth,trim=0 0 1.3cm 0.5cm, clip]{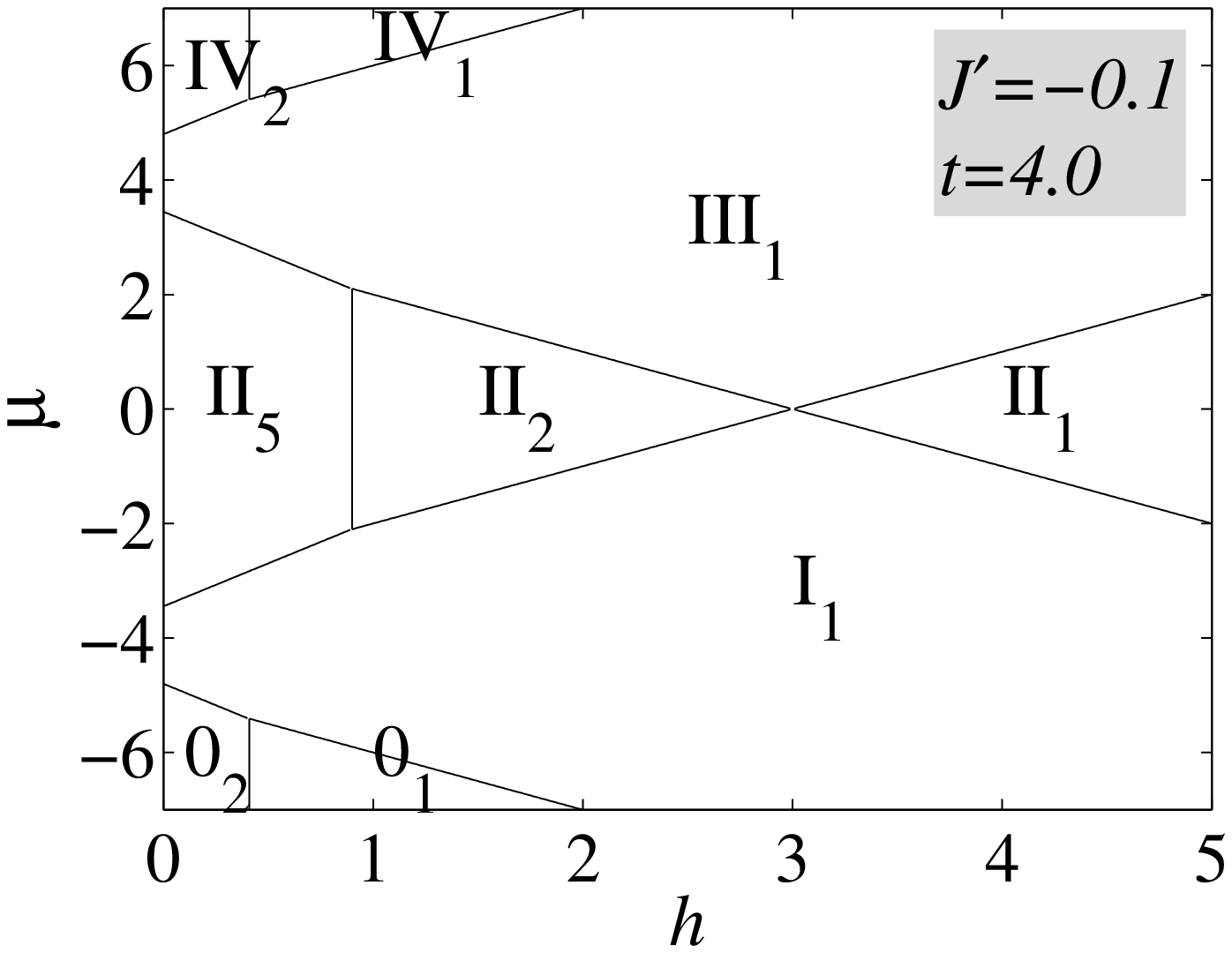}
\includegraphics[width=0.3\textwidth,trim=0 0 1.3cm 0.5cm, clip]{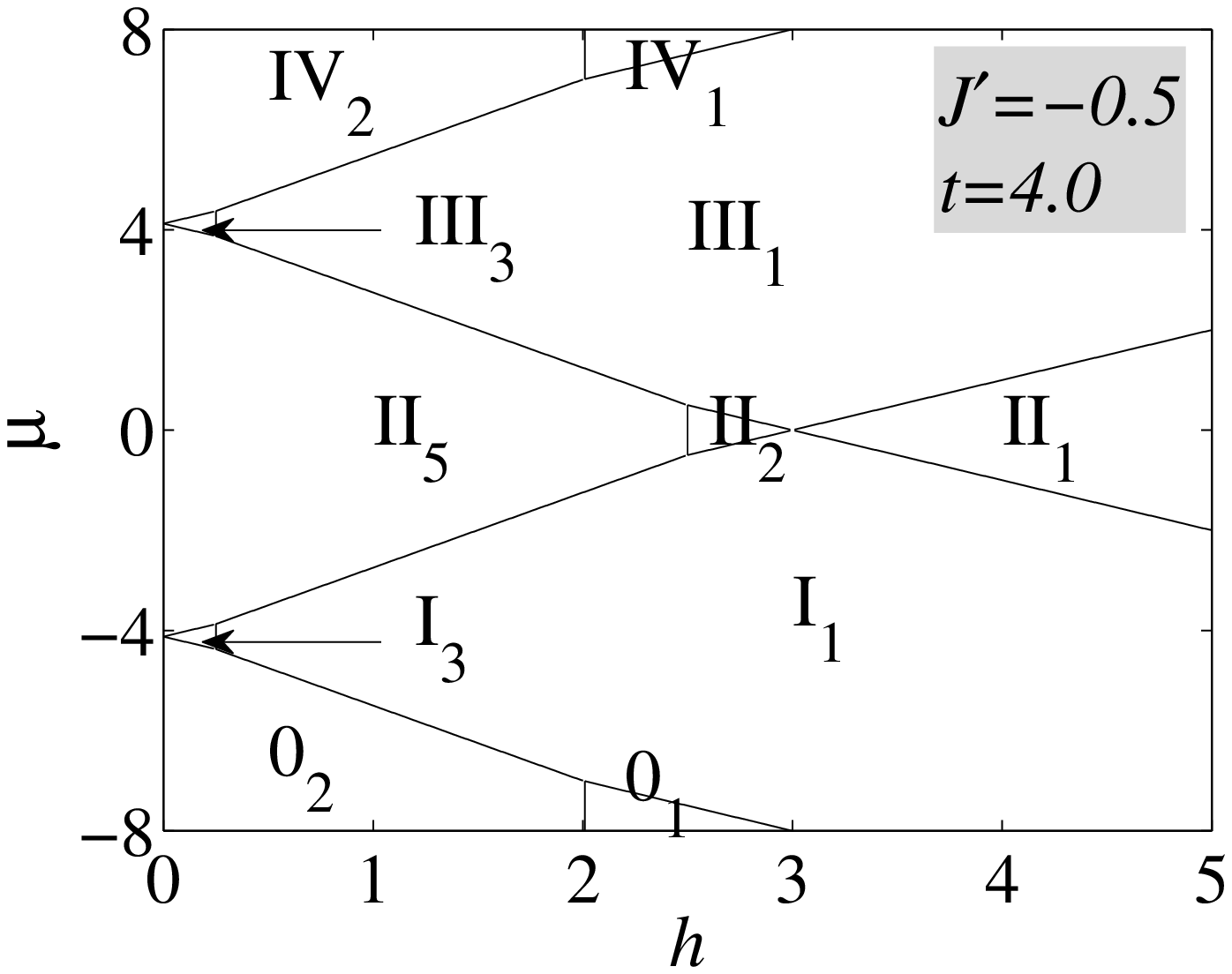}
\hspace{0.3\textwidth}
\caption{\small Ground-state phase diagrams in the $\mu$-$h$ plane for $J=1$ and selected values of $J'\leq 0$ and $t$.}
\label{fig3}
\end{center}
\end{figure*}

\begin{figure}
 \centering
 {\includegraphics[width=0.45\textwidth,trim=0 0 1.3cm 0.5cm, clip]{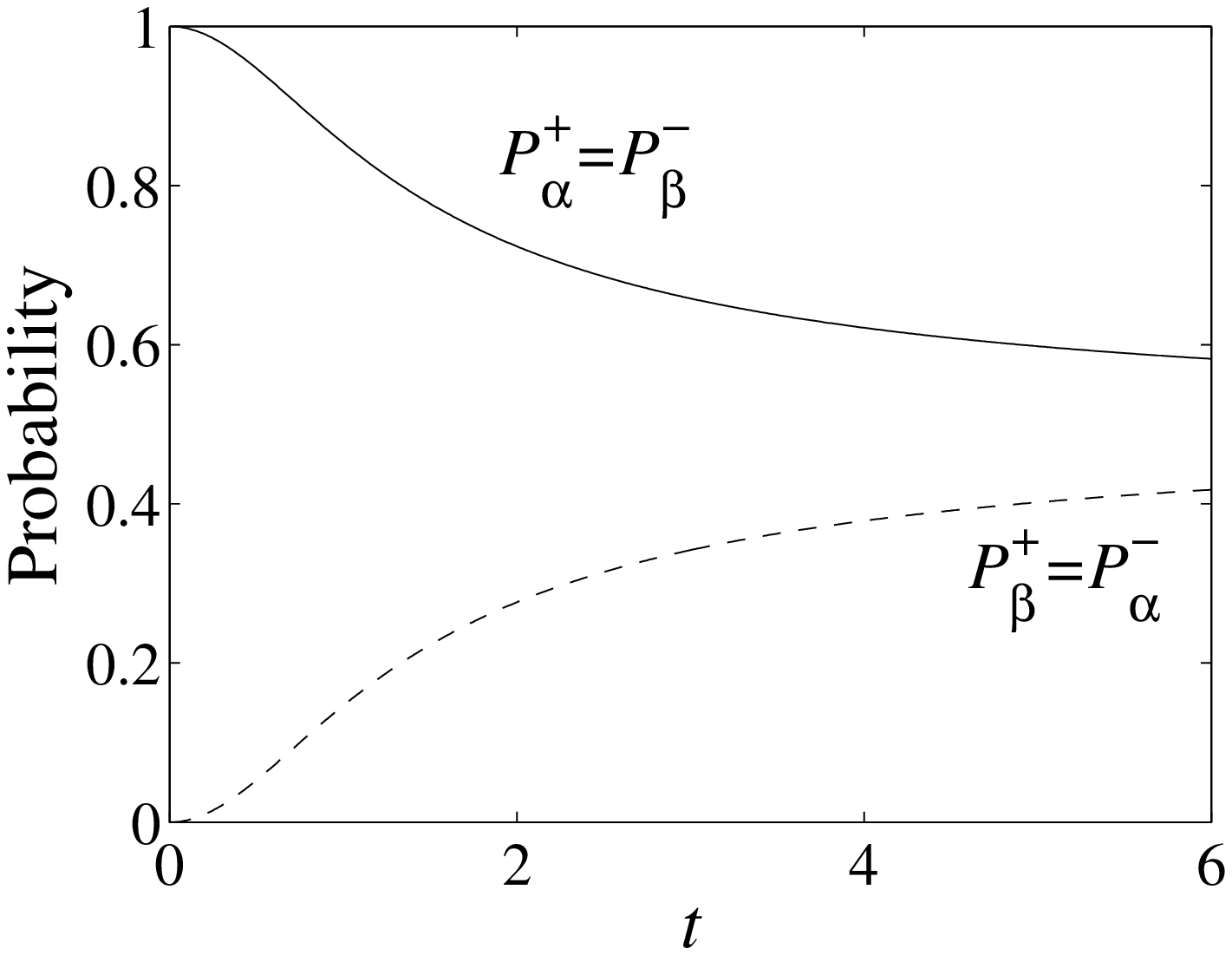}}
  \begin{tikzpicture}[overlay]
       \node at (-4.8,3.3) {\includegraphics[scale=0.28]{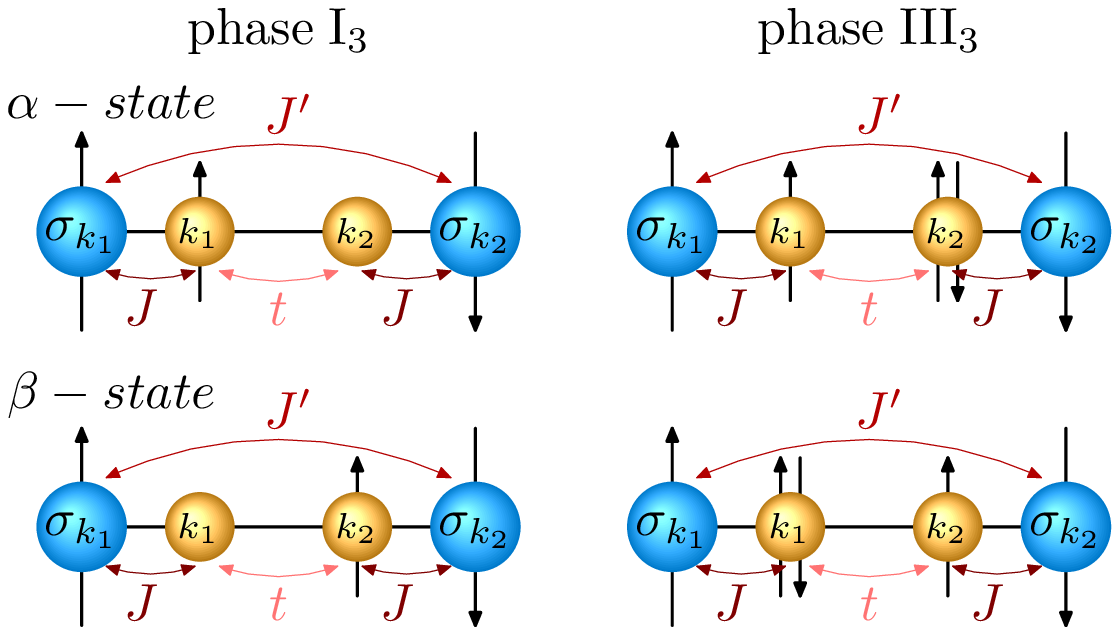}};
  \end{tikzpicture}
\caption{\small The occurrence probabilities of microstates within the ground states I$_3$ and III$_3$ (see Tab.~\ref{tab1}), where $P^{\pm}_{\alpha}$ determines the probability of the microstate $|\uparrow,0\rangle_k$ with the corresponding probability amplitude $\alpha$  and $P^{\pm}_{\beta}$ stands for the probability of the microstate $|0,\uparrow\rangle_k$ with the corresponding probability amplitude $\beta$. The upper index determines sign of the coupling constant $J$, i.e., '+' for $J>0$ and '-' for $J<0$. Inset: the microstates entering a quantum superposition within the phase I$_3$ and III$_3$ with probability amplitudes $\alpha$ and $\beta$, respectively.}
\label{fig3a}
\end{figure}

Contrary to this, the AF interaction $J'<0$ may significantly influence the former phase diagrams and generate novel magnetic phases, especially, at low magnetic fields. Since the driving force of their existence originates from the $AF$ interaction $J'<0$, naturally, the stability of novel phases arises as a response to strengthening of the AF interaction $|J'|$ (see Fig.~\ref{fig3}). All new phases are indeed characterized by the AF arrangement of magnetic moments in the spin subsystem (see Tab.~\ref{tab1}), in accordance with the AF character of the interaction $J'<0$. In the parameter space, where the effect of coupling constant $J$ is negligible ($n_k=0$ or $n_k=4$), the spin order is strictly determined by the competition between the field term $h$ and the coupling constant $J'<0$, while the value of hopping term $t$ becomes unimportant. The same conclusion can be reached for 
the phase boundaries 0$_1$-0$_2$ or IV$_1$-IV$_2$ emerging at $h/q=-J'$. By contrast, the increasing hopping term $t$ has a significant effect on the  electron subsystem within the novel phase I$_3$/III$_3$ and, thus, it dramatically changes the stability of these phases. However, it can be understood from Fig.~\ref{fig3a} that the hopping process of the mobile electrons effectively decouples the localized spins  within the novel phases I$_3$ or III$_3$ in the limit of sufficiently strong hopping term $t$ what is in sharp contrast  to the phases I$_1$/III$_1$ with the F alignment of the localized spins.
Similarly to the case $J'>0$, the AF further-neighbor interaction $J'<0$ strongly affects presence/existence of field-driven phase transitions. Contrary to the former case, the increasing $|J'|$ shifts the phase boundary to the higher magnetic fields with exception  of the phase boundary between II$_1$-II$_2$ phases. It is noteworthy that sufficiently strong value of $J'$ can fully suppress presence of the phase II$_2$ for a strong electron correlation ($t=4$) and thus, it can reduce the number of field-driven phase transitions.  For completeness, let us quote analytical expressions  for remaining phase boundaries occurring in the phase diagrams for $J>0$ and $J'<0$, as derived from a comparison of the  energies given in Tab.~\ref{tab1}.
\begin{eqnarray}
\hspace{-0.2cm}
\begin{array}{ll}
{\rm 0}_2{\rm -I}_1\;/\;{\rm III}_1{\rm -IV}_2:&\!\!\!\! \mu\!=\!(-1)^{u}(-J-h-t-2J'-2h/q), \\
{\rm 0}_2{\rm -I}_2\;/\;{\rm III}_2{\rm -IV}_2:&\!\!\!\!\mu\!=\!(-1)^{u}(J-t-h-2J'+2h/q), \\
{\rm I}_3{\rm -II}_1\;/\;{\rm II}_1{\rm -III}_3:&\!\!\!\!\mu\!=\!(-1)^{u}\left(-2(J+J'+h/q)-h\right.\\
&+\sqrt{J^2+t^2}),\\
{\rm I}_3{\rm -II}_2\;/\;{\rm II}_2{\rm -III}_3:&\!\!\!\! \mu\!=\!(-1)^{u}\left(-2(J'+t+h/q)+h\right.\\
&+\sqrt{J^2+t^2}), \\
{\rm I}_3{\rm -II}_5\;/\;{\rm II}_5{\rm -III}_3:&\!\!\!\! \mu\!=\!(-1)^{u}(h-\sqrt{J^2+t^2}),\\
{\rm I}_1{\rm -I}_3\;/\;{\rm III}_3{\rm -III}_1:&\!\!\!\! 2h/q\!=\!-J-2J'-t+\sqrt{J^2+t^2}.
\end{array} 
\hspace*{-1cm}\label{eqA4}
\end{eqnarray}
\\
\subsection{The antiferromagnetic case $J<0$}
\begin{figure*}[t!]
\begin{center}
\includegraphics[width=0.3\textwidth,trim=0 0 1.3cm 0.5cm, clip]{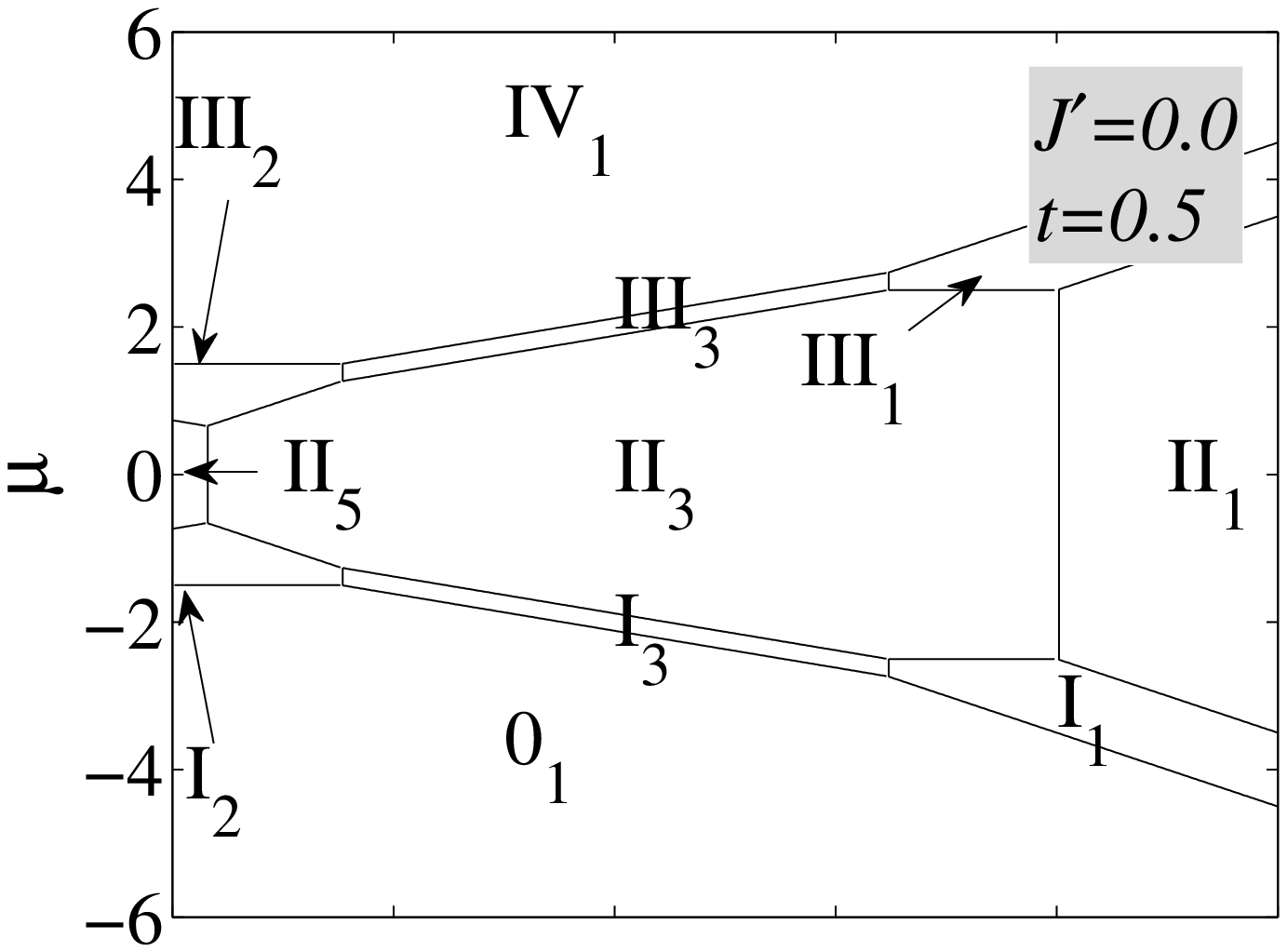}
\includegraphics[width=0.3\textwidth,trim=0 0 1.3cm 0.5cm, clip]{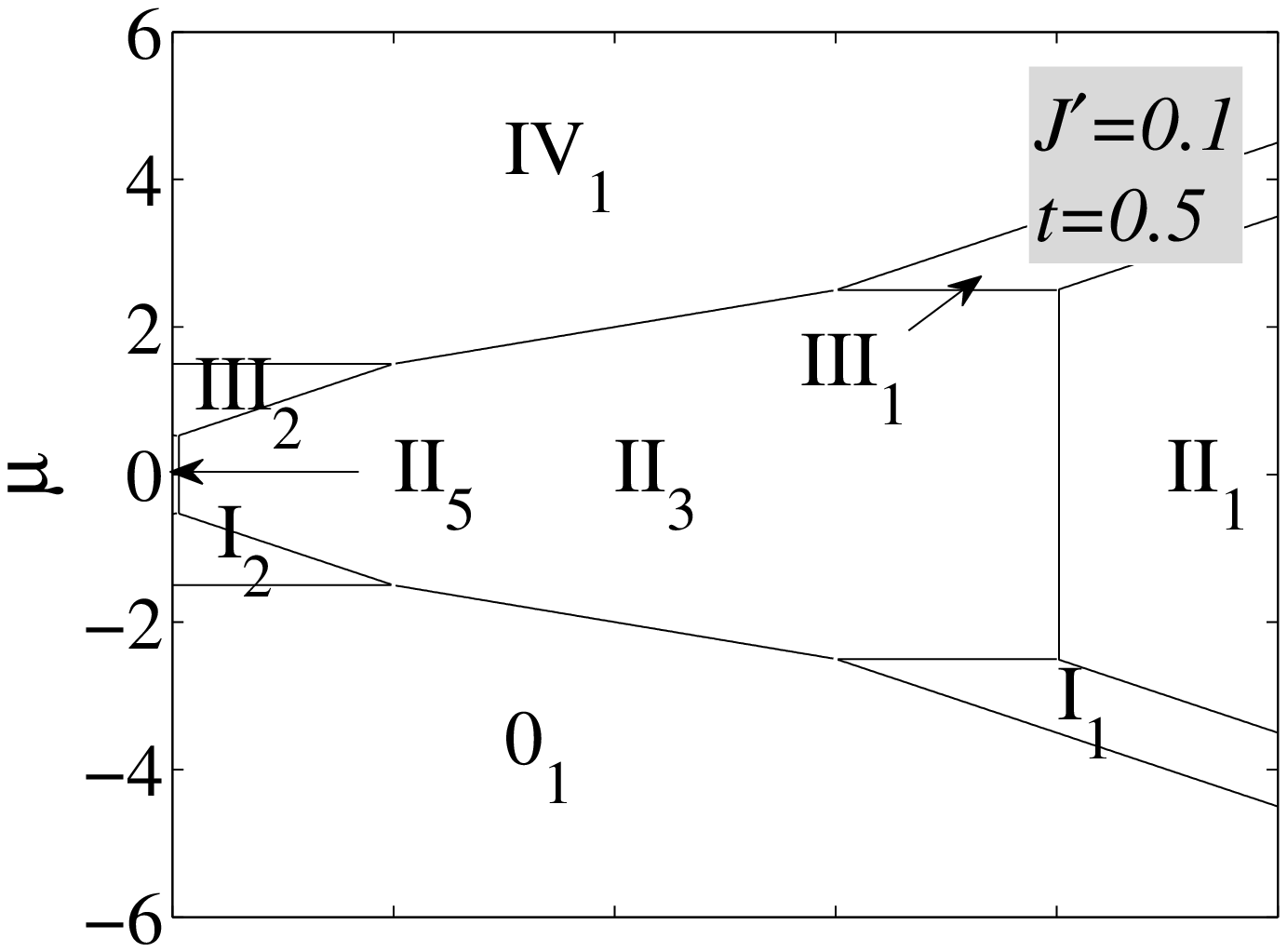}
\includegraphics[width=0.3\textwidth,trim=0 0 1.3cm 0.5cm, clip]{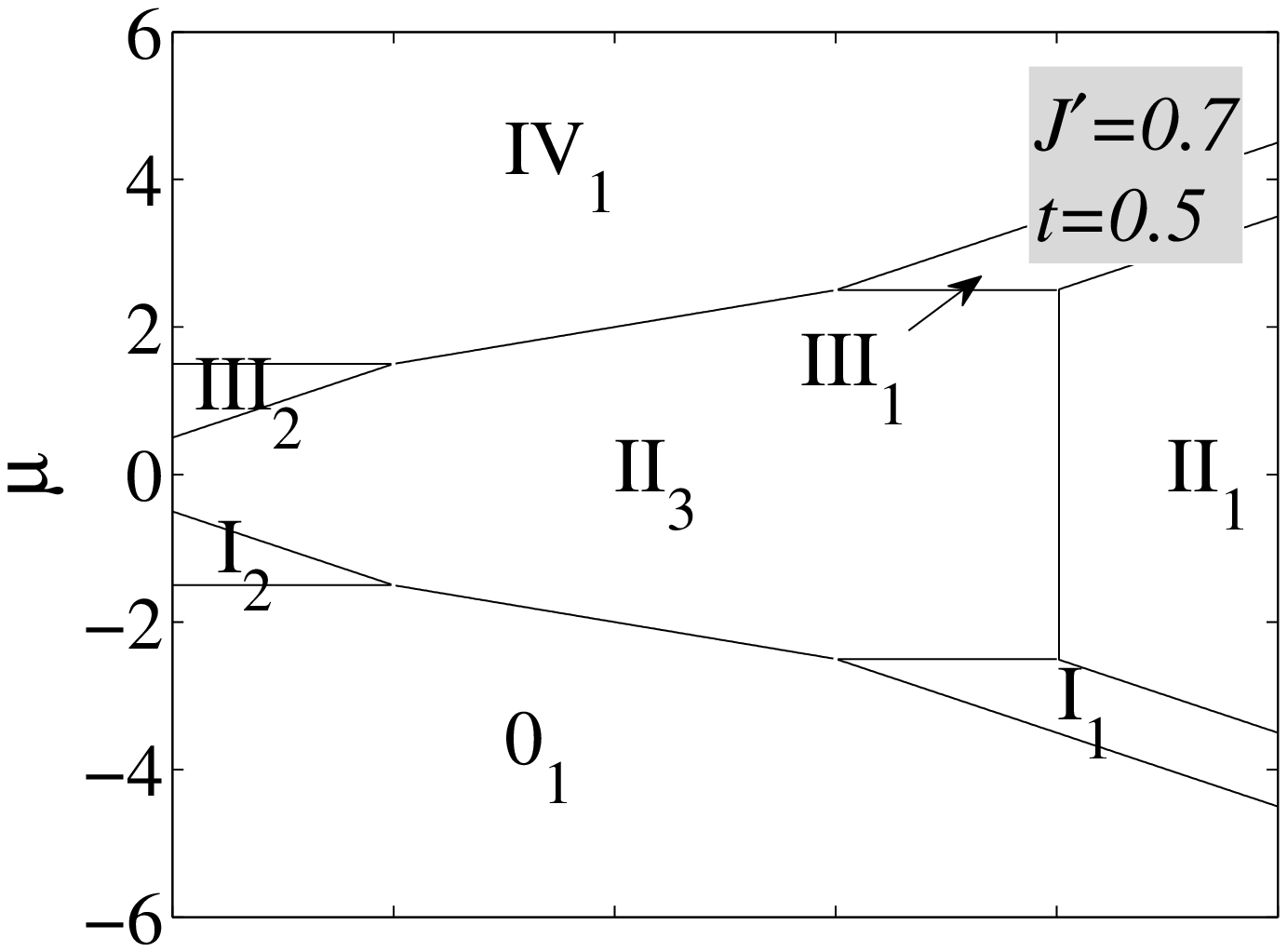}\\
\vspace*{-0.39cm}
\includegraphics[width=0.3\textwidth,trim=0 0 1.3cm 0.5cm, clip]{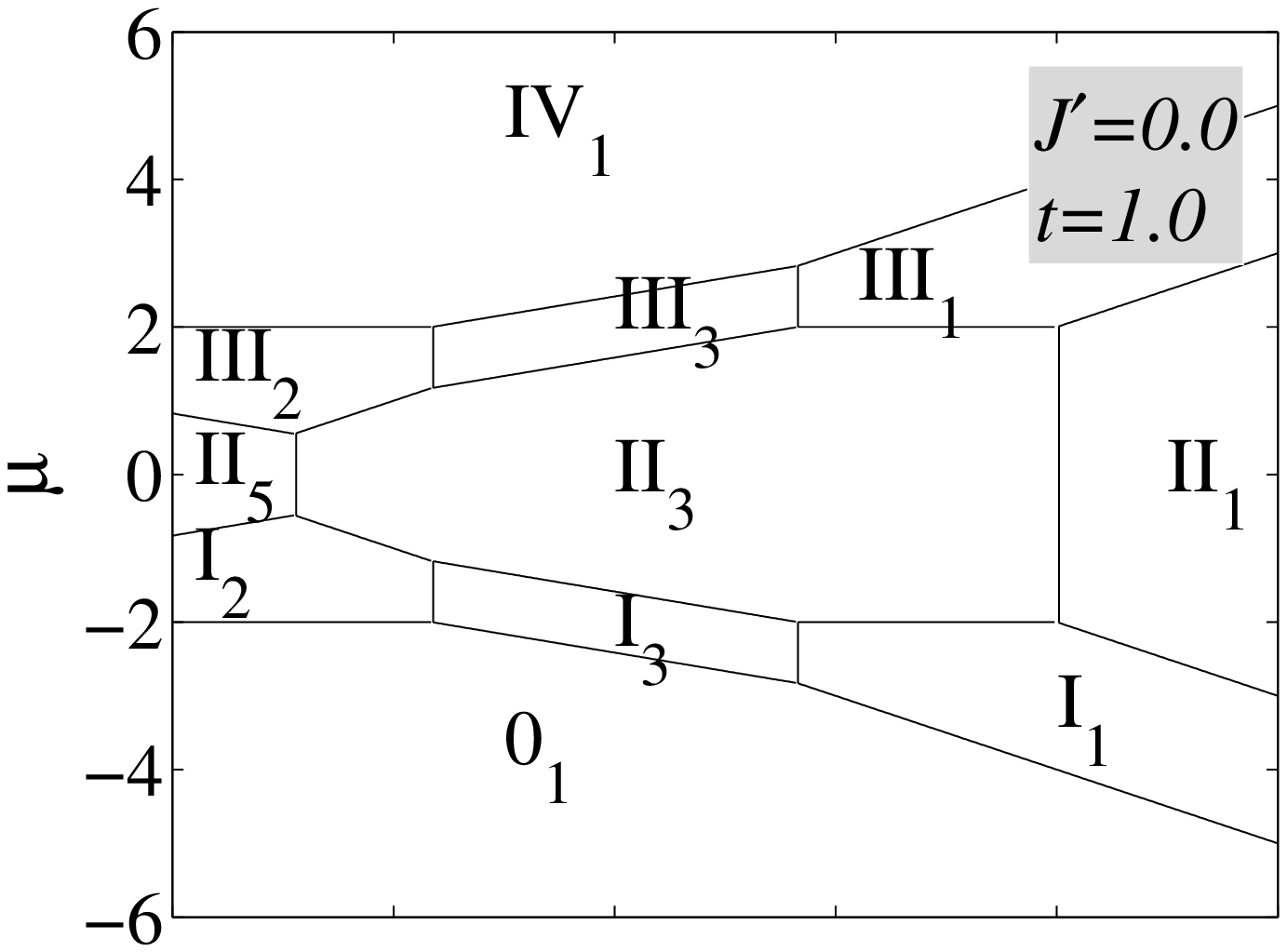}
\includegraphics[width=0.3\textwidth,trim=0 0 1.3cm 0.5cm, clip]{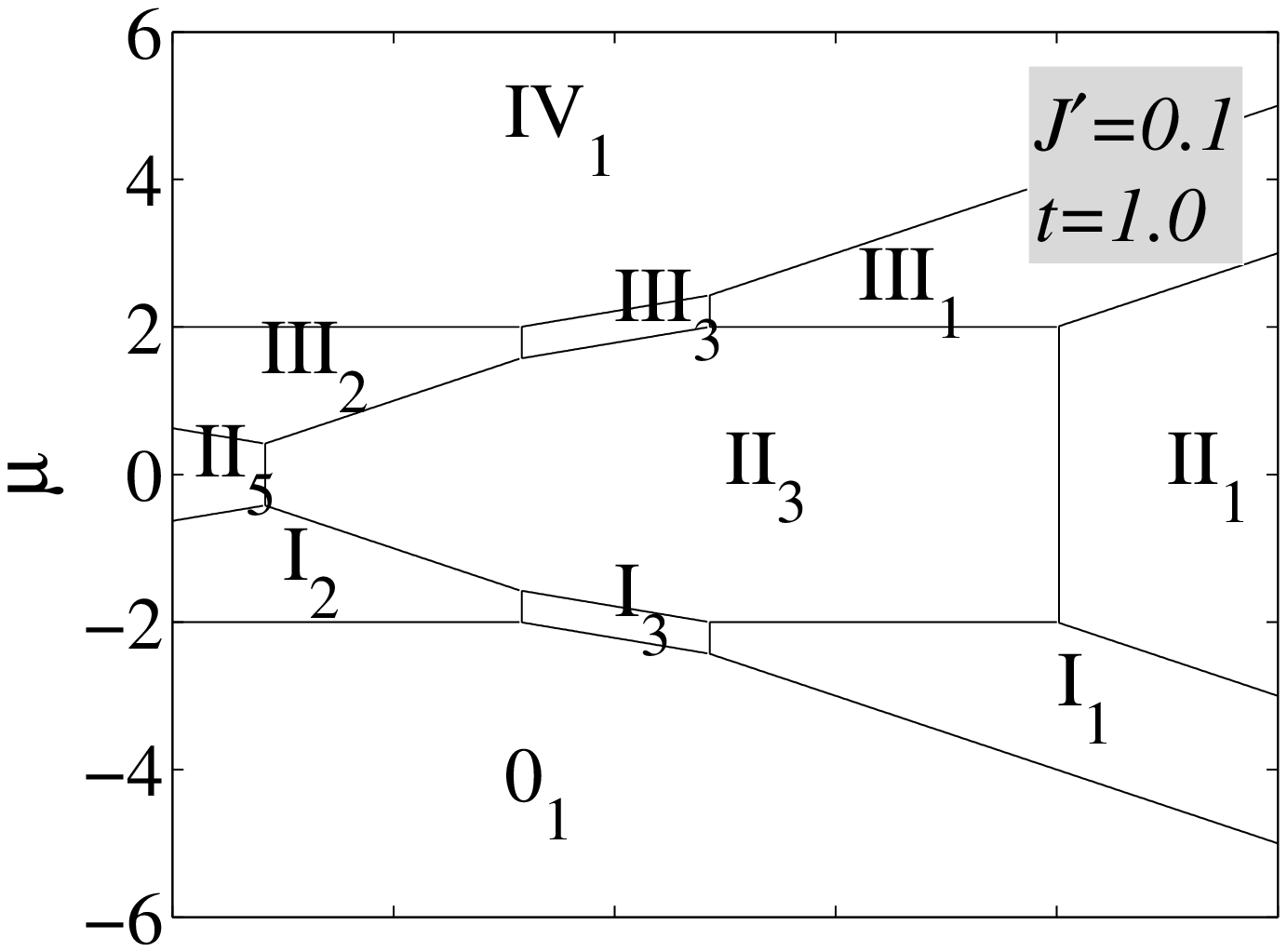}
\includegraphics[width=0.3\textwidth,trim=0 0 1.3cm 0.5cm, clip]{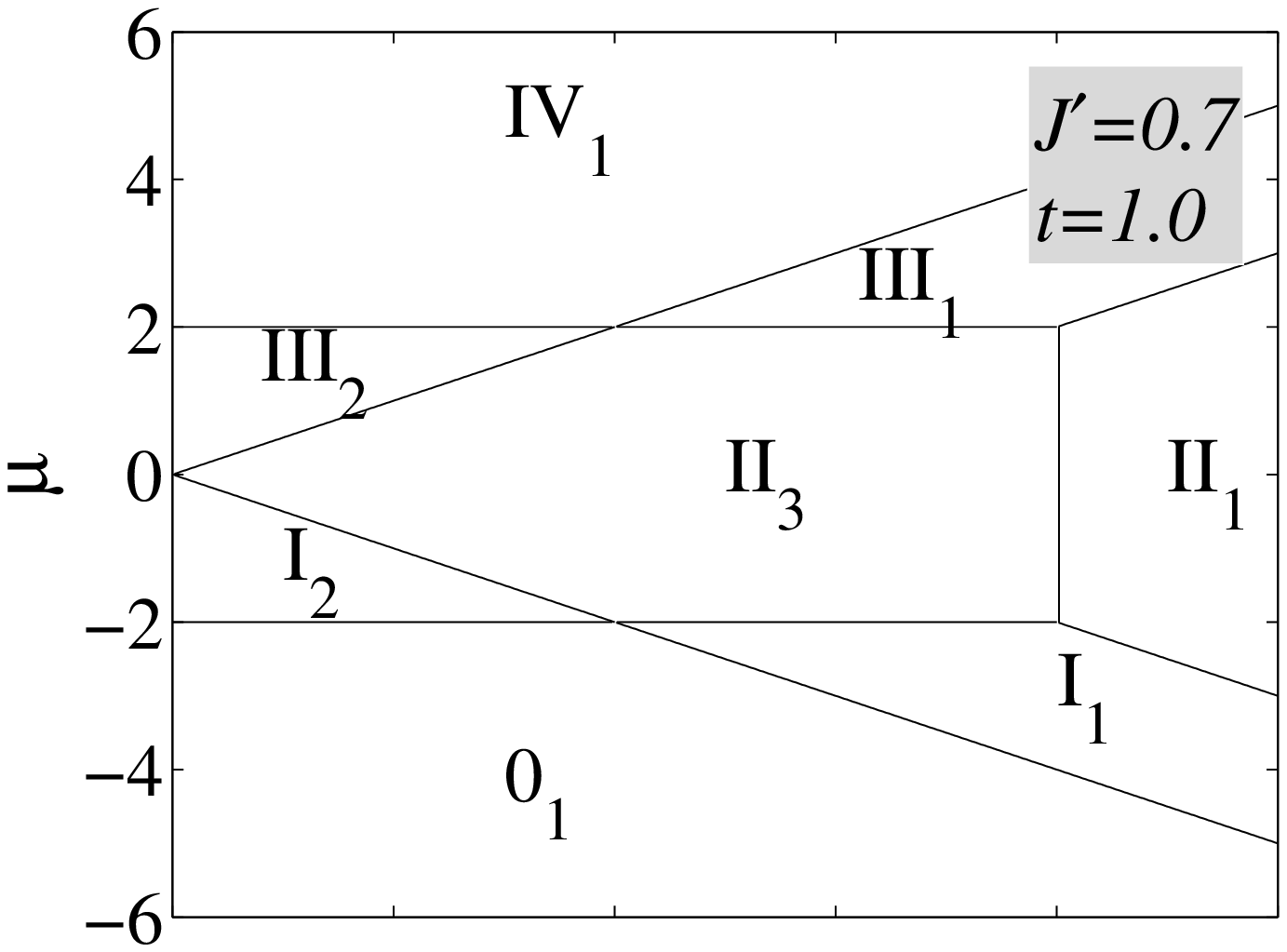}\\
\vspace*{-0.39cm}
\includegraphics[width=0.3\textwidth,trim=0 0 1.3cm 0.5cm, clip]{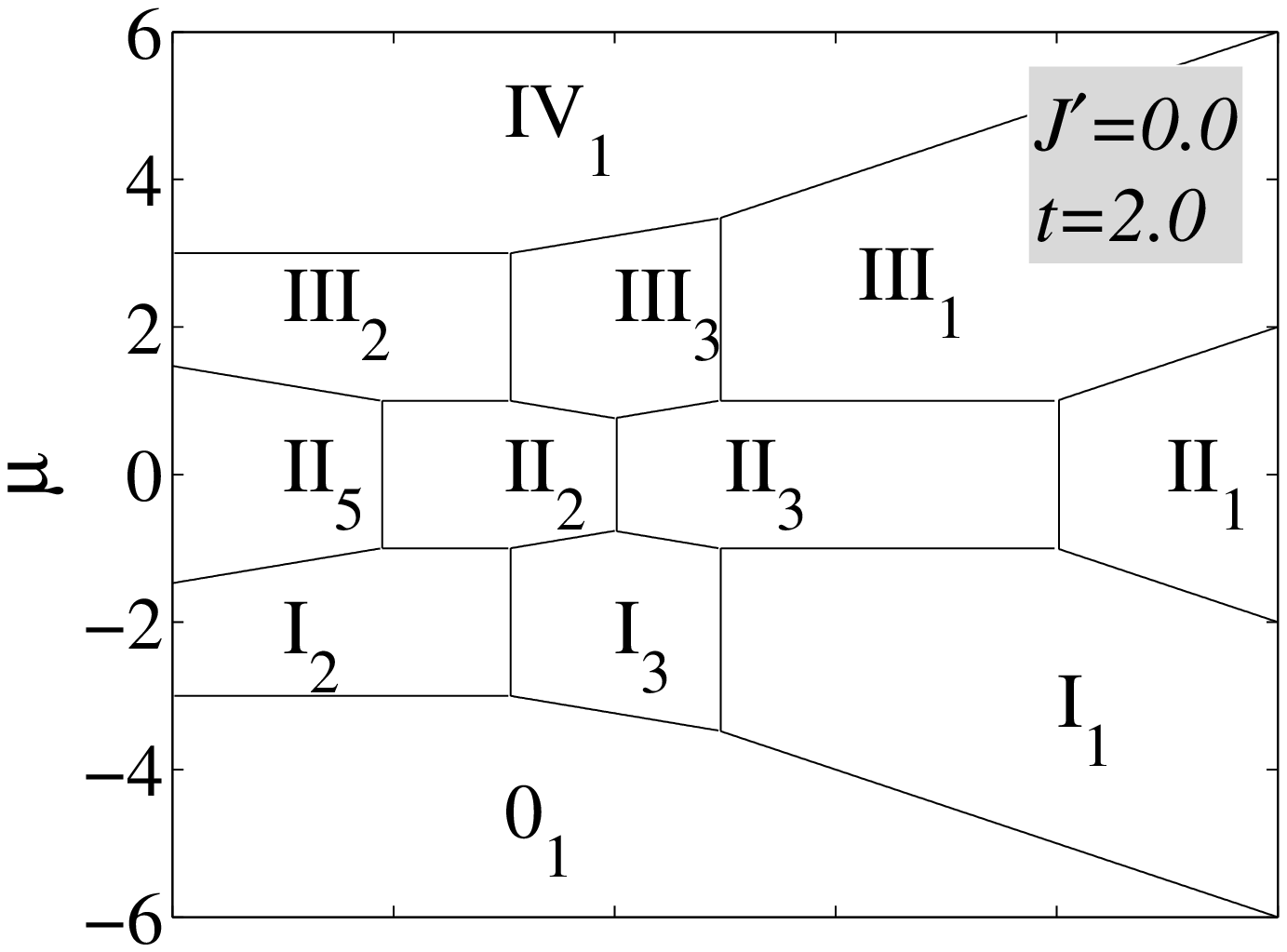}
\includegraphics[width=0.3\textwidth,trim=0 0 1.3cm 0.5cm, clip]{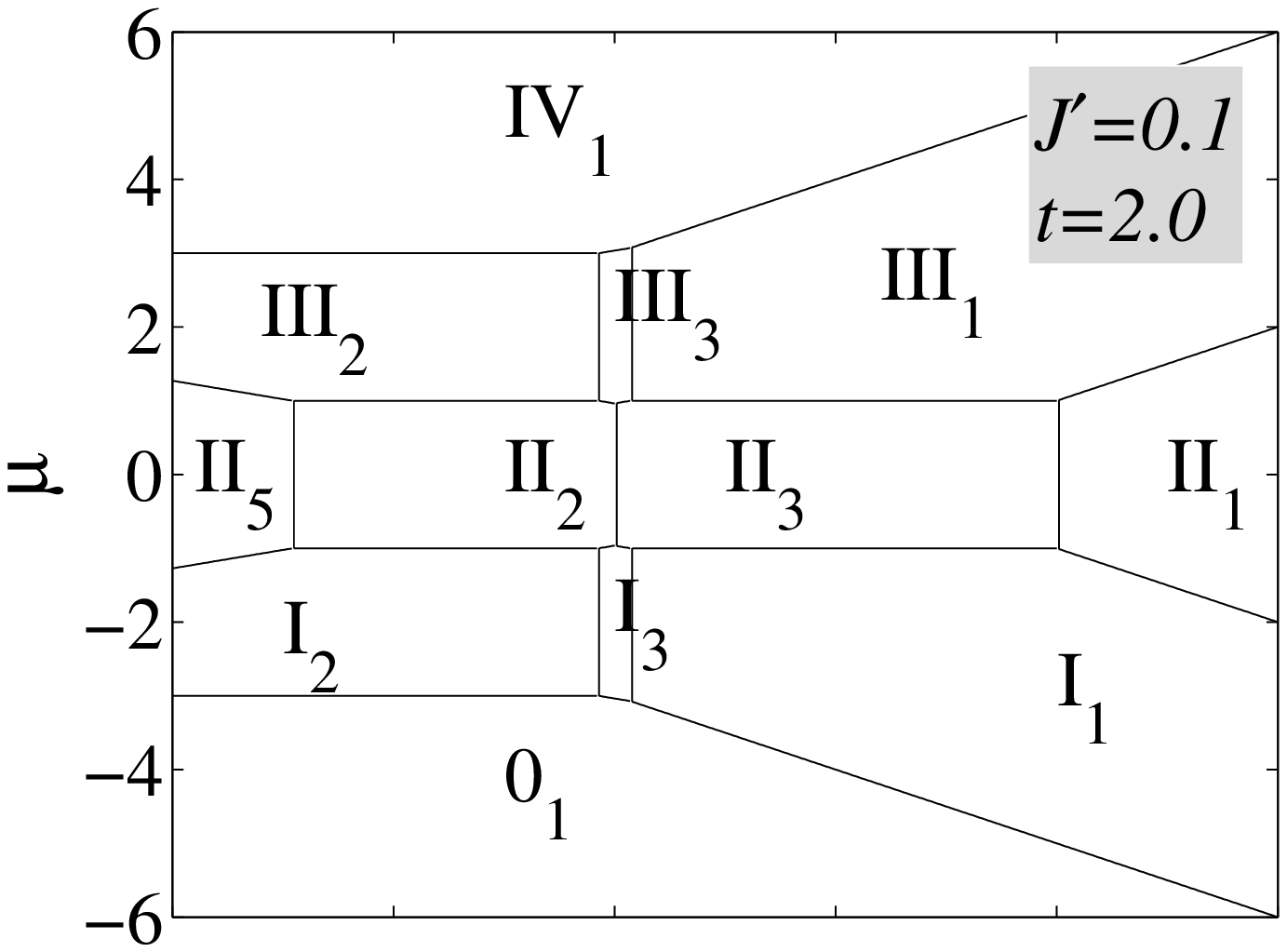}
\includegraphics[width=0.3\textwidth,trim=0 0 1.3cm 0.5cm, clip]{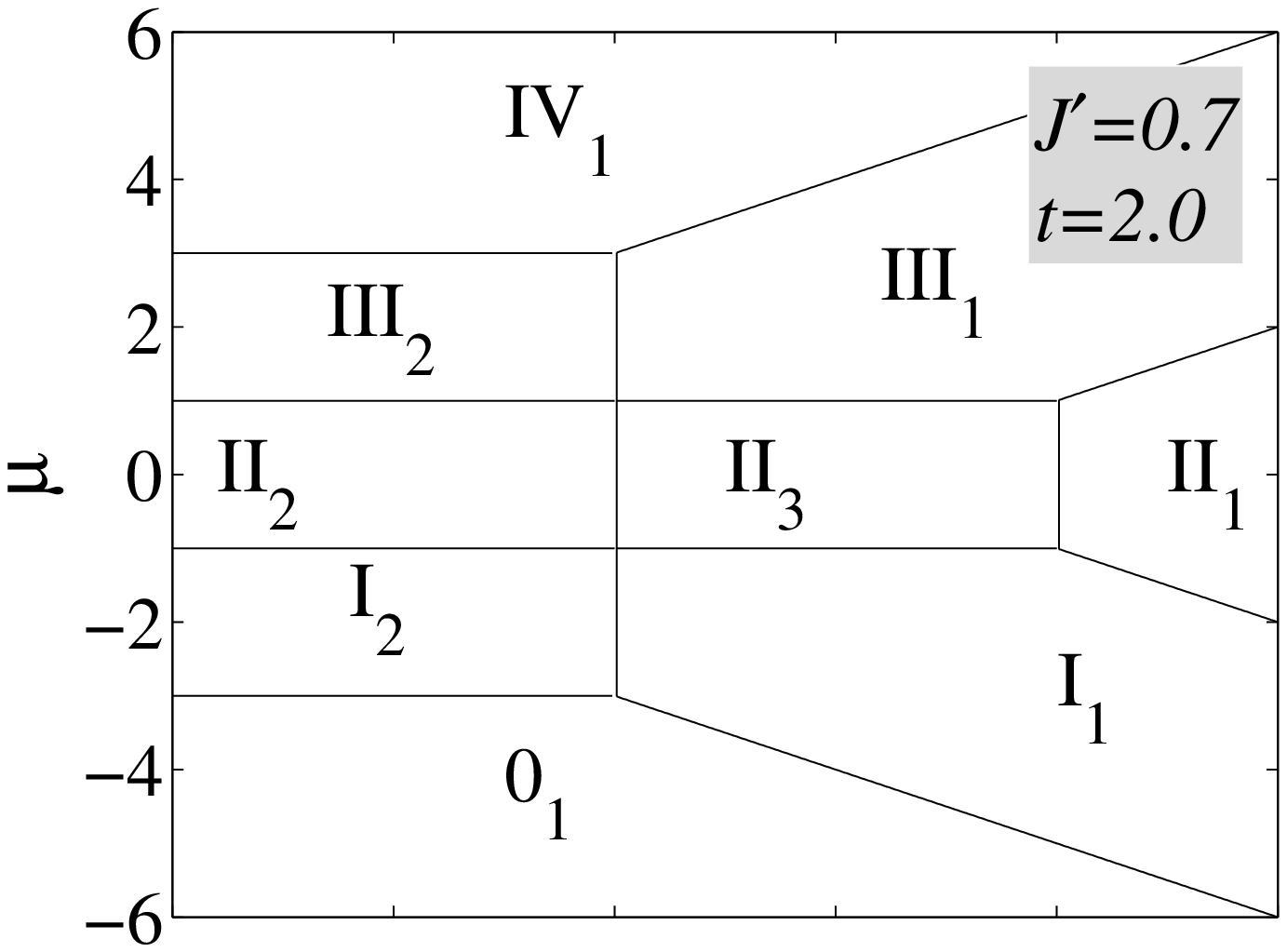}\\
\vspace*{-0.39cm}
\includegraphics[width=0.3\textwidth,trim=0 0 1.3cm 0.5cm, clip]{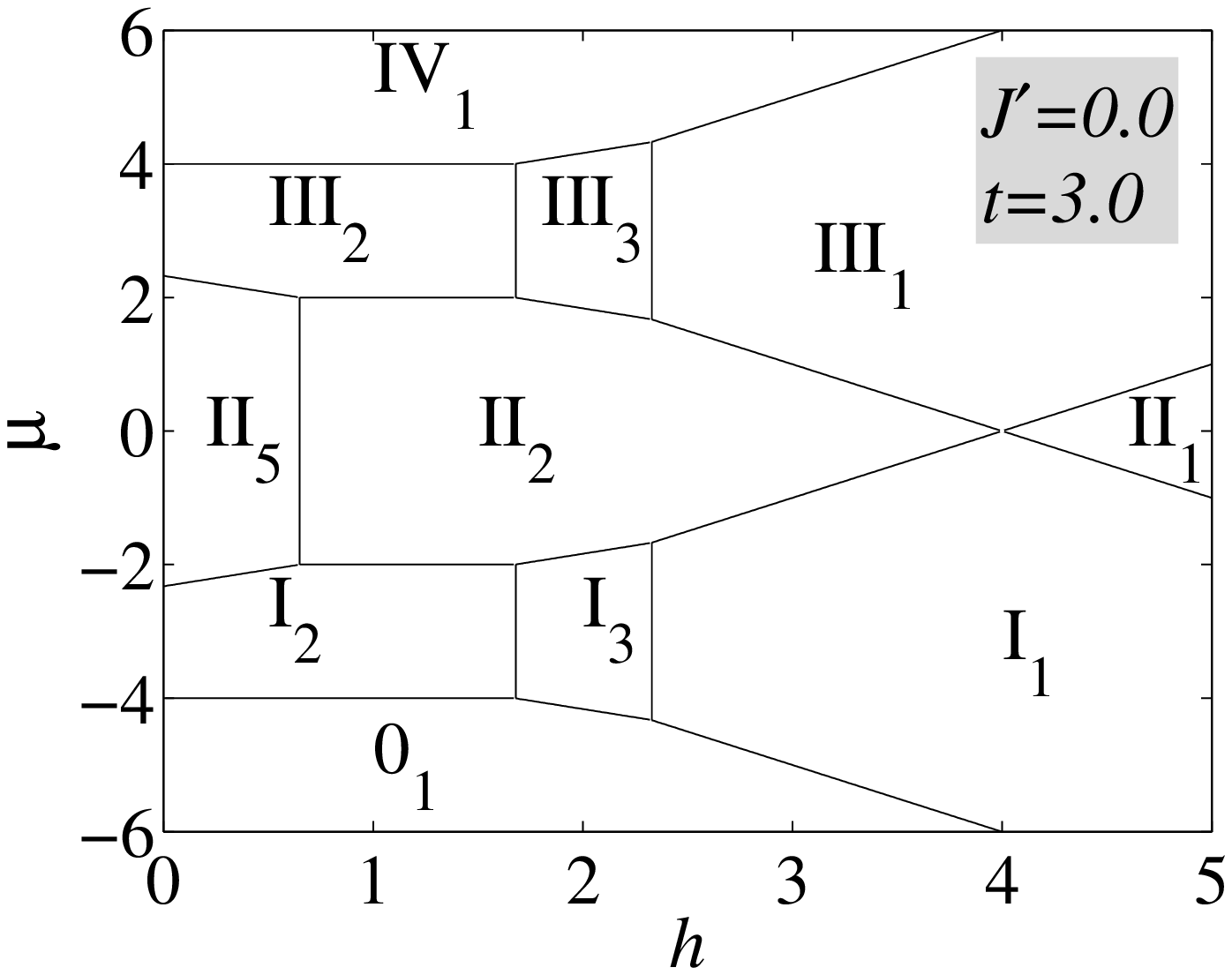}
\includegraphics[width=0.3\textwidth,trim=0 0 1.3cm 0.5cm, clip]{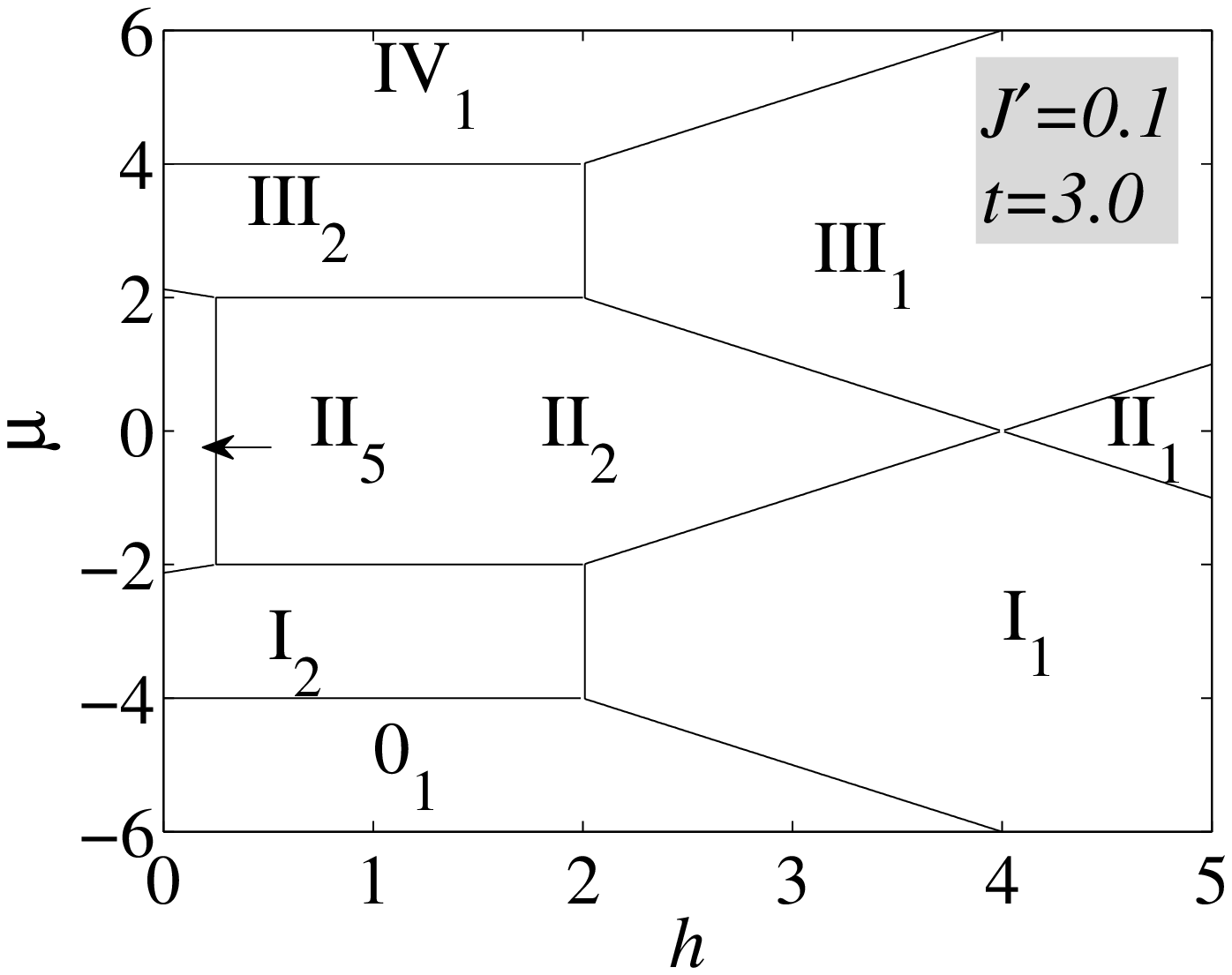}
\includegraphics[width=0.3\textwidth,trim=0 0 1.3cm 0.5cm, clip]{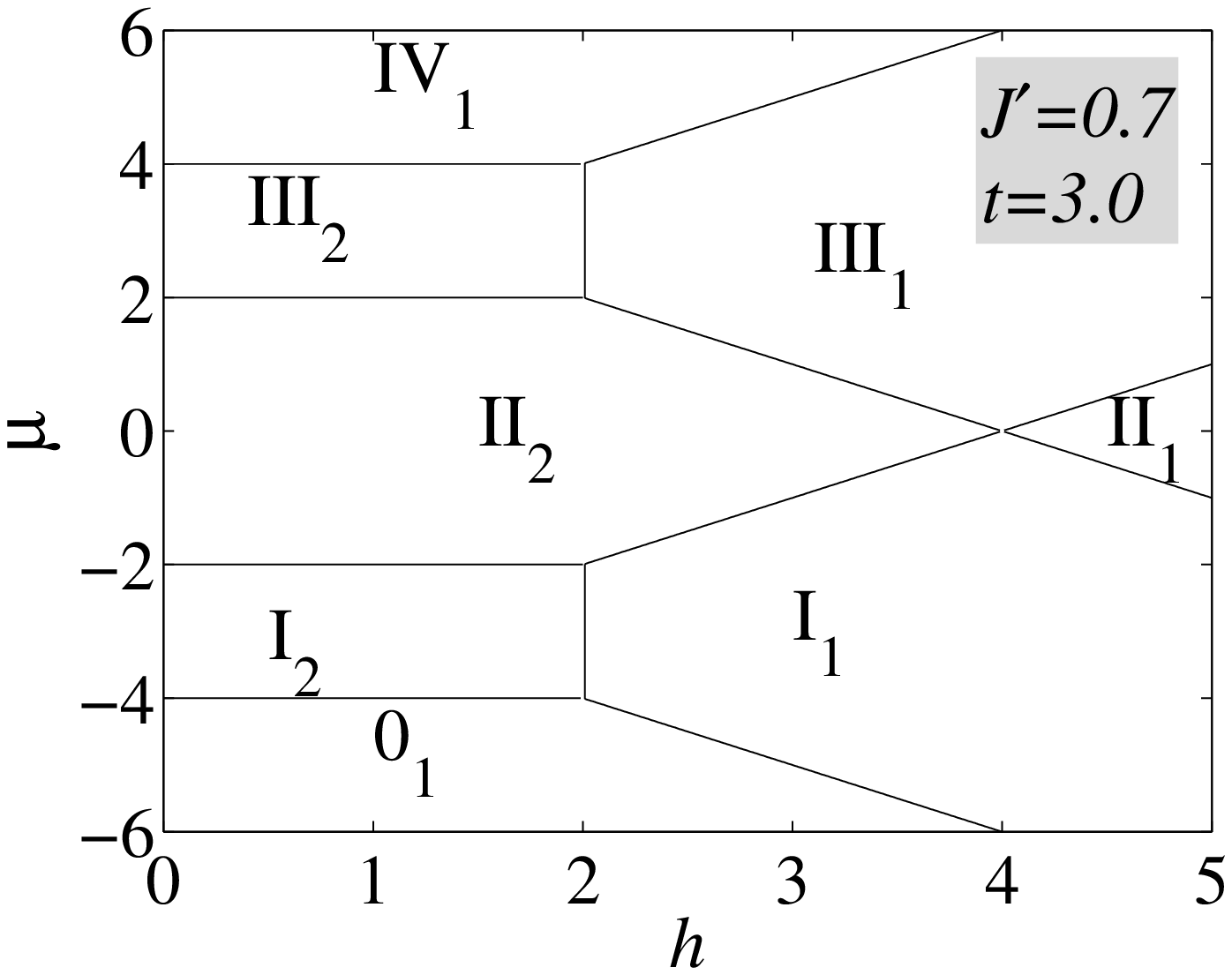}
\caption{\small Ground-state phase diagrams in the $\mu$-$h$ plane for  $J=1$ and selected values of $J'\geq 0$ and $t$.}
\label{fig4}
\end{center}
\end{figure*}

The situation for the AF coupling $J<0$ is more complicated. Under this condition, the spin subsystem in the case of the fully occupied or empty electron counterpart is always oriented ferromagnetically, since the effect of magnetic field dominates over all other  forces. In an uncompensated electron limit ($n_k=1$ or $n_k=3$), the AF coupling $J<0$ generates at low magnetic fields new phases I$_2$ and III$_2$ with a different magnetism in both subsystems. Since the effect of applied field is smaller in comparison with the AF coupling $J<0$, the localized spins are aligned in opposite to the magnetic field in order to preserve the AF character of the spin-electron coupling $J<0$. Naturally, the increment of external field leads to a suppression  of the AF coupling $J<0$  and the F character becomes dominant. This transition is consequently realized  through the intermediate phases I$_3$ and III$_3$ to the final F phases I$_1$ and III$_1$, as evidenced by two field-driven phase transitions. The first transition between the phases   I$_2$-I$_3$ (III$_2$-III$_3$) depends on all model parameters, $2h/q=-J+2J'+t-\sqrt{J^2+t^2}$, and it is shifted to the higher fields as the hopping term $t$ increases. The second transition also depends on all model parameters, but the increasing hopping term $t$ shifts its occurrence inversely. Consequently, both transitions can merge together into the $t$-invariant phase boundary, $h/q=-J/2$, between the phases I$_1$-I$_2$ or III$_1$-III$_2$ for a sufficiently large hopping term $t$. It should be mentioned that the occurrence probabilities of two electron microstates in the phase I$_3$ or III$_3$ evolve inversely with respect to the $J>0$ case (Fig.~\ref{fig3a}). Furthermore,  the AF coupling $J<0$ produces another novel magnetic phase II$_3$ located at a half filling. This phase is also characterized by a different magnetism of both subsystems due to the AF character of the spin-electron coupling $J<0$. Surprisingly, the phase II$_3$ occurs at relatively high magnetic fields in contrast to the phases with the AF ordering in one (II$_2$) or both (II$_5$) subsystems emergent in a low-field region. It can be seen from  Fig.~\ref{fig4} that the system exhibits field-driven phase transitions also at half filling in the limit with absence of spin-spin interaction $J'=0$, where their number can be tuned by the hopping term $t$. The rigorous expressions for three of them are given by Eq.~(\ref{eqA2}), while the remaining two field-driven phase transitions occur at:
\begin{eqnarray}
\begin{array}{ll}
{\rm II}_1{\rm -II}_3:&\!\!\!\! h/q\!=\!-J,\\
{\rm II}_3{\rm -II}_5:&\!\!\!\! h/q\!=\!\left(J-J'+\sqrt{J^2+t^2}\right)/(q-1).
\end{array}
\label{eqA6} 
\end{eqnarray}
Let us turn our attention to the effect of the further-neighbor  interaction $J'$ on the ground-state properties. As could be expected, the F interaction $J'>0$ stabilizes the spontaneous F ordering and reduces the AF ones, whereas the phases emerging for the AF coupling $J<0$ dominate in weak magnetic fields. In the strong-field limit the phases emerging for the F coupling $J>0$ become dominant. Obviously, the non-zero F spin-spin interaction $J'>0$ in combination with the AF spin-electron one $J<0$ is not able to generate a novel magnetic phase, but it only favors some of existing phases at the expense of others. The most interesting finding in this parameter space is the fact that all field-induced phase transitions occurring in the phase diagrams depend only on values of $h$ and $J$ for sufficiently large $J'$ as well as $t$. By contrast, the AF interaction $J'<0$ may generate novel phases 0$_2$ and IV$_2$ at low magnetic fields and it also generates the novel phase II$_4$ at relatively high magnetic fields at the half-filled band case on assumption that the hopping term is sufficiently strong. In this phase the AF coupling $J'<0$ between the localized spins is sufficient to  preserve their antiparallel  orientation, but the  magnetic field is strong enough to align the electron subsystem into the direction of external magnetic field. 
\begin{figure*}[t!]
\begin{center}
\includegraphics[width=0.3\textwidth,trim=0 0 1.3cm 0.5cm, clip]{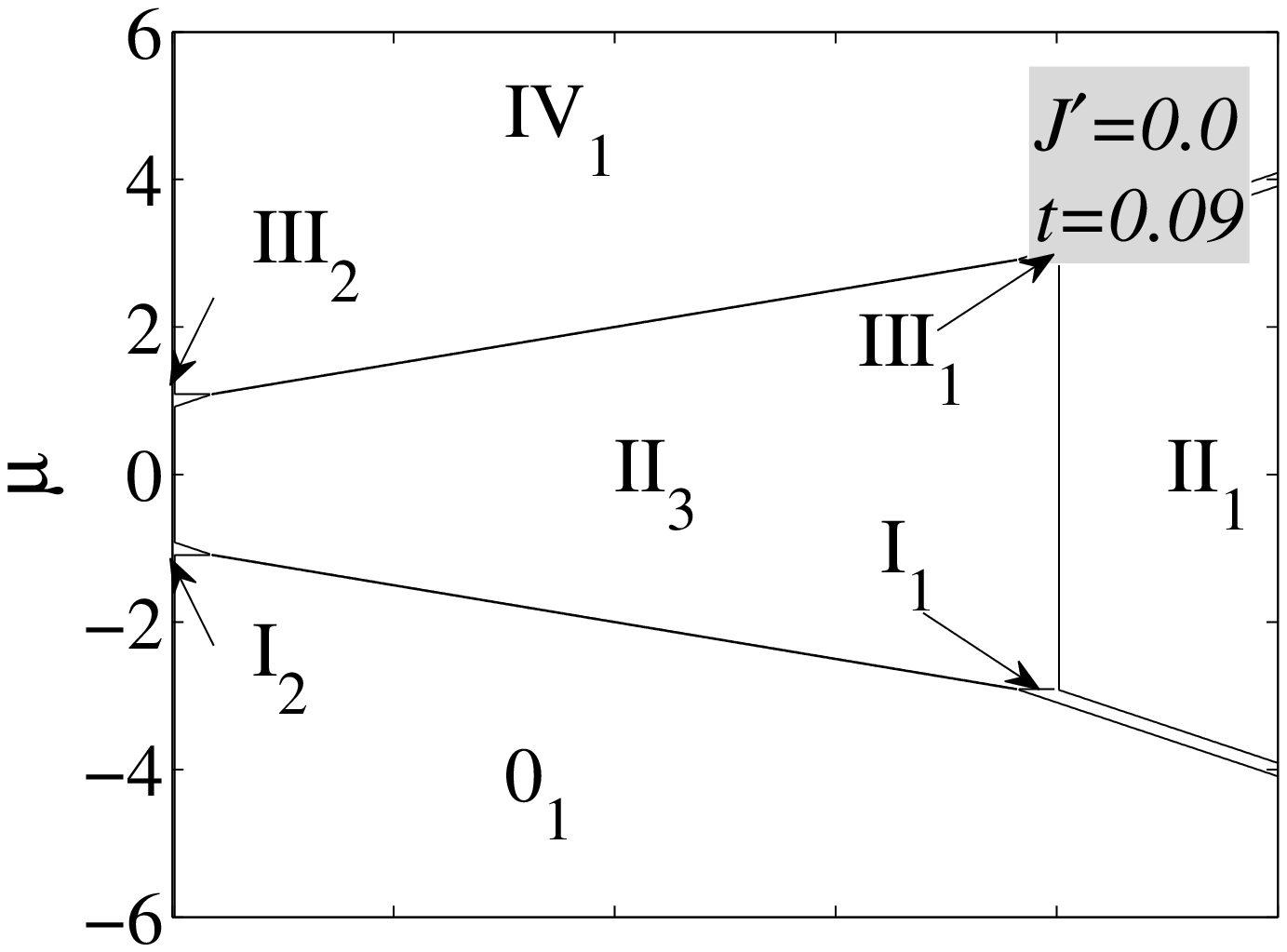}
\includegraphics[width=0.3\textwidth,trim=0 0 1.3cm 0.5cm, clip]{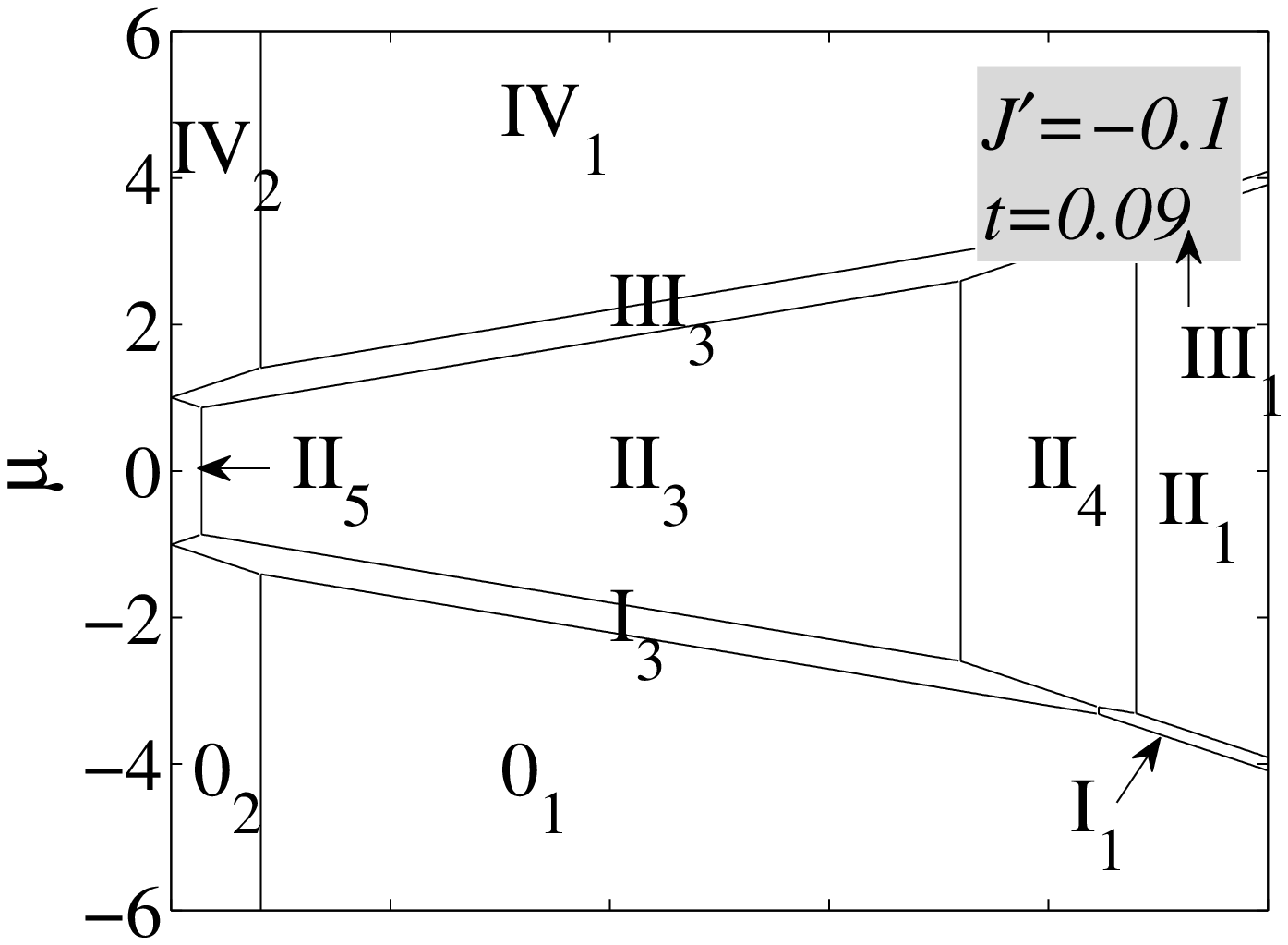}
\includegraphics[width=0.3\textwidth,trim=0 0 1.3cm 0.5cm, clip]{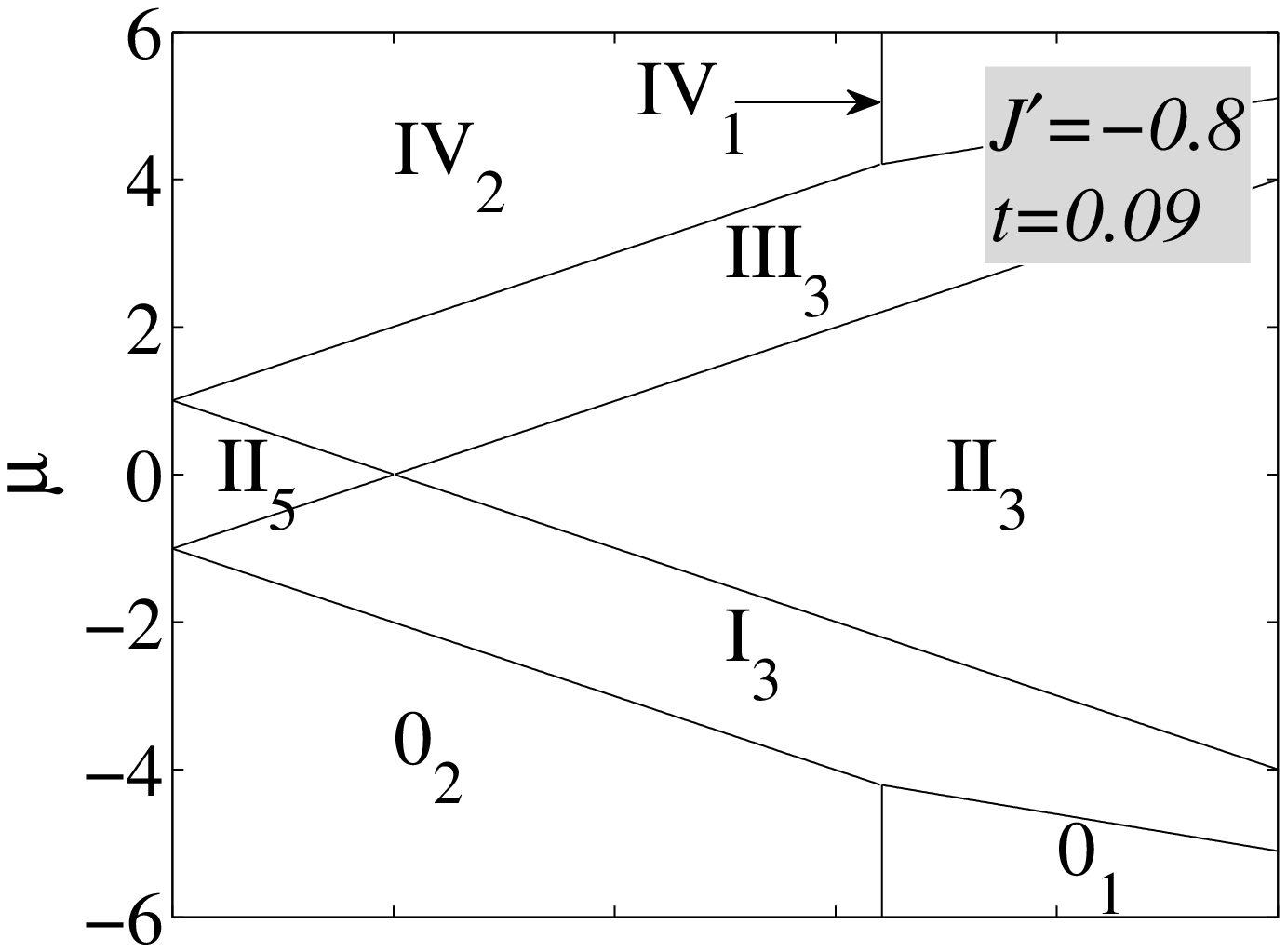}
\\\vspace*{-0.39cm}
\includegraphics[width=0.3\textwidth,trim=0 0 1.3cm 0.5cm, clip]{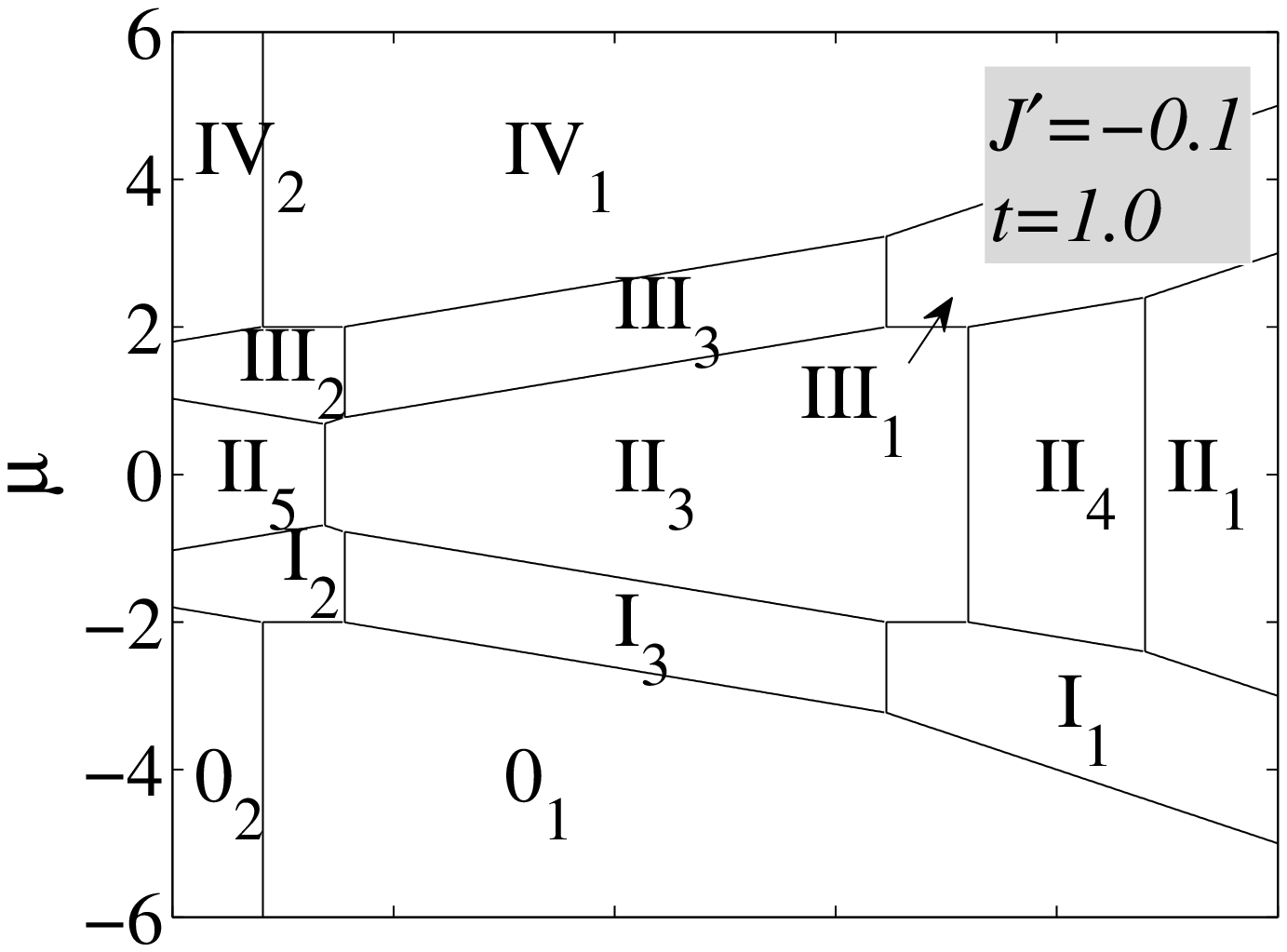}
\includegraphics[width=0.3\textwidth,trim=0 0 1.3cm 0.5cm, clip]{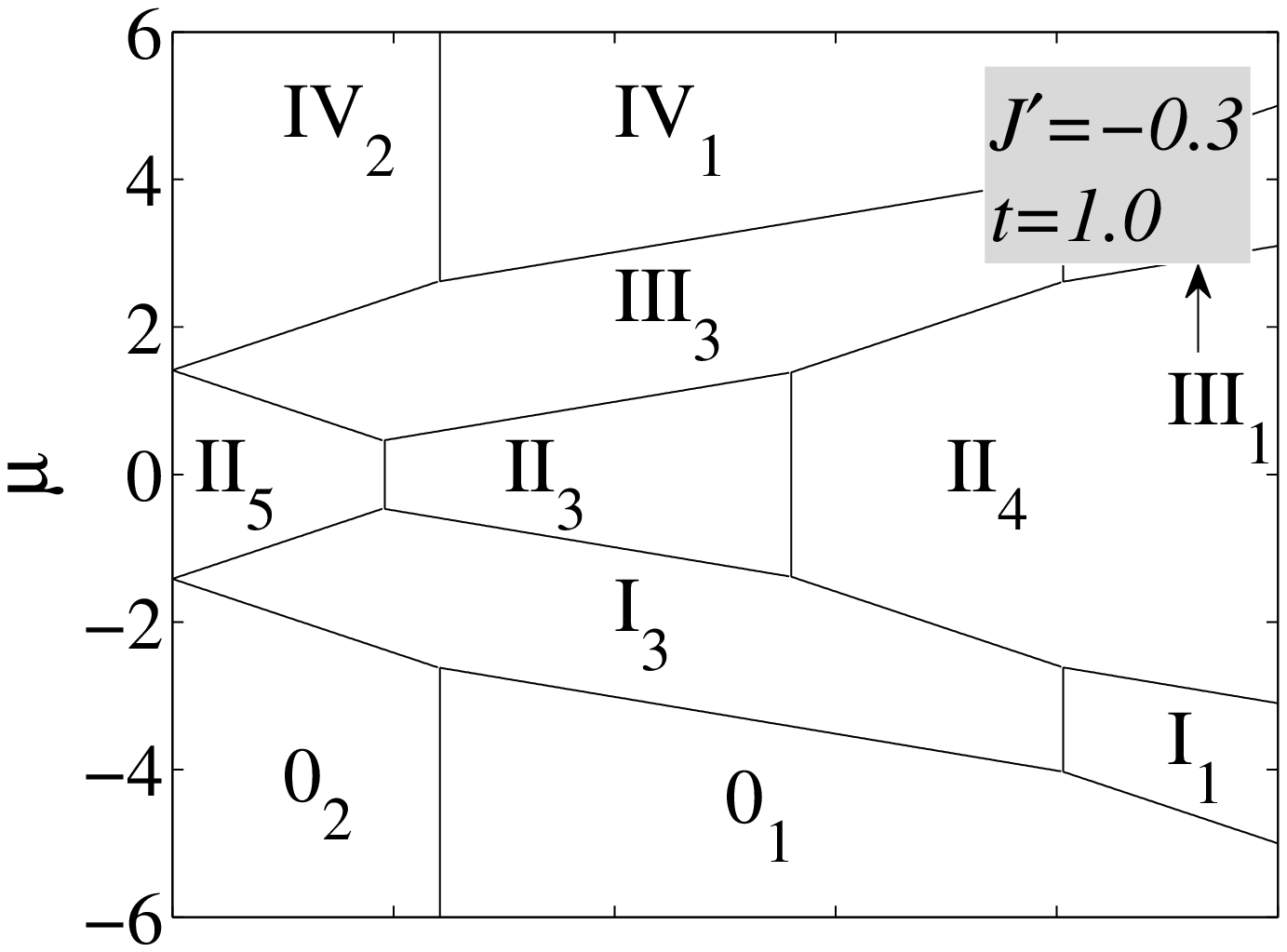}
\includegraphics[width=0.3\textwidth,trim=0 0 1.3cm 0.5cm, clip]{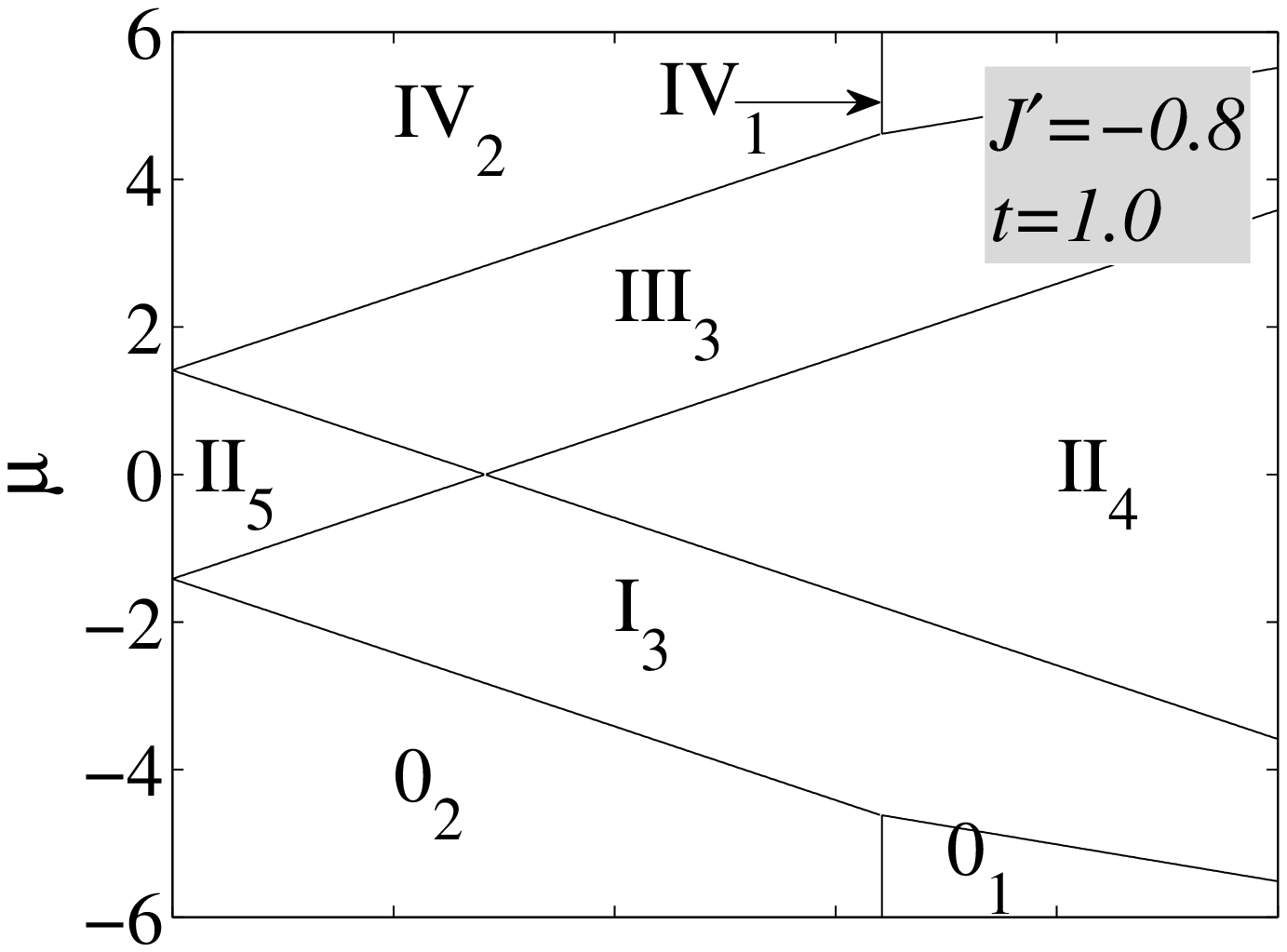}\\
\vspace*{-0.39cm}
\includegraphics[width=0.3\textwidth,trim=0 0 1.3cm 0.5cm, clip]{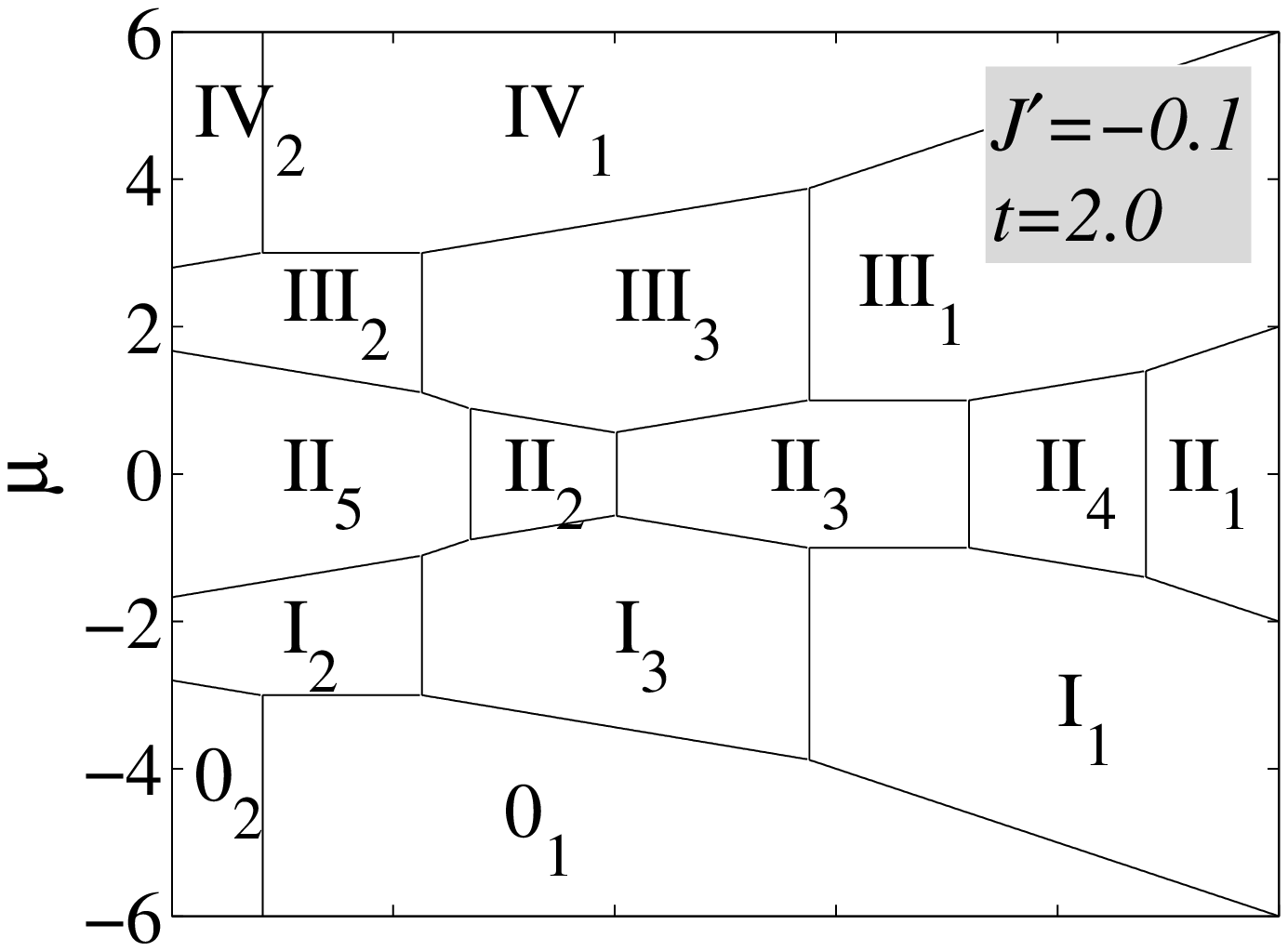}
\includegraphics[width=0.3\textwidth,trim=0 0 1.3cm 0.5cm, clip]{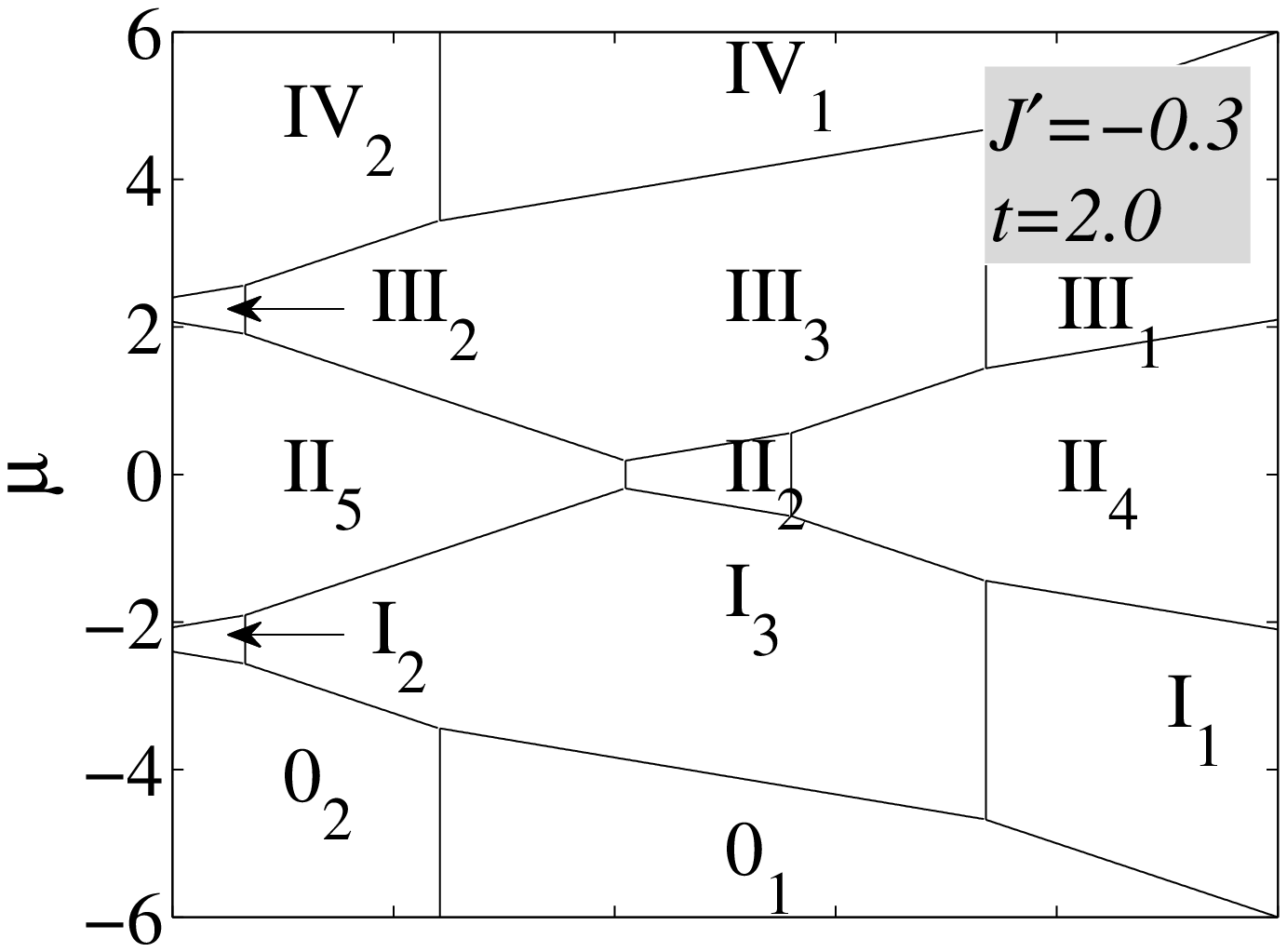}
\includegraphics[width=0.3\textwidth,trim=0 0 1.3cm 0.5cm, clip]{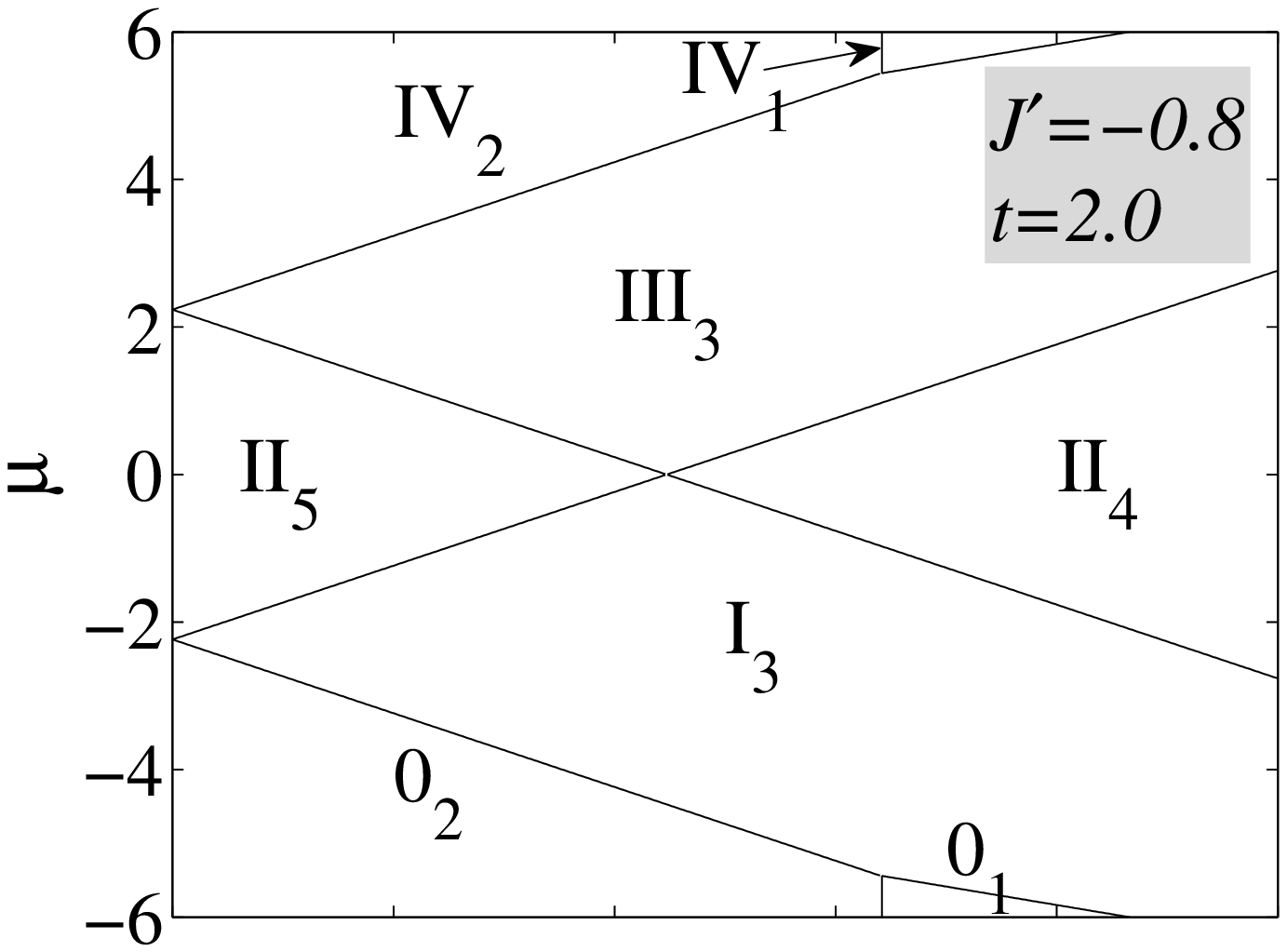}\\
\vspace*{-0.39cm}
\includegraphics[width=0.3\textwidth,trim=0 0 1.3cm 0.5cm, clip]{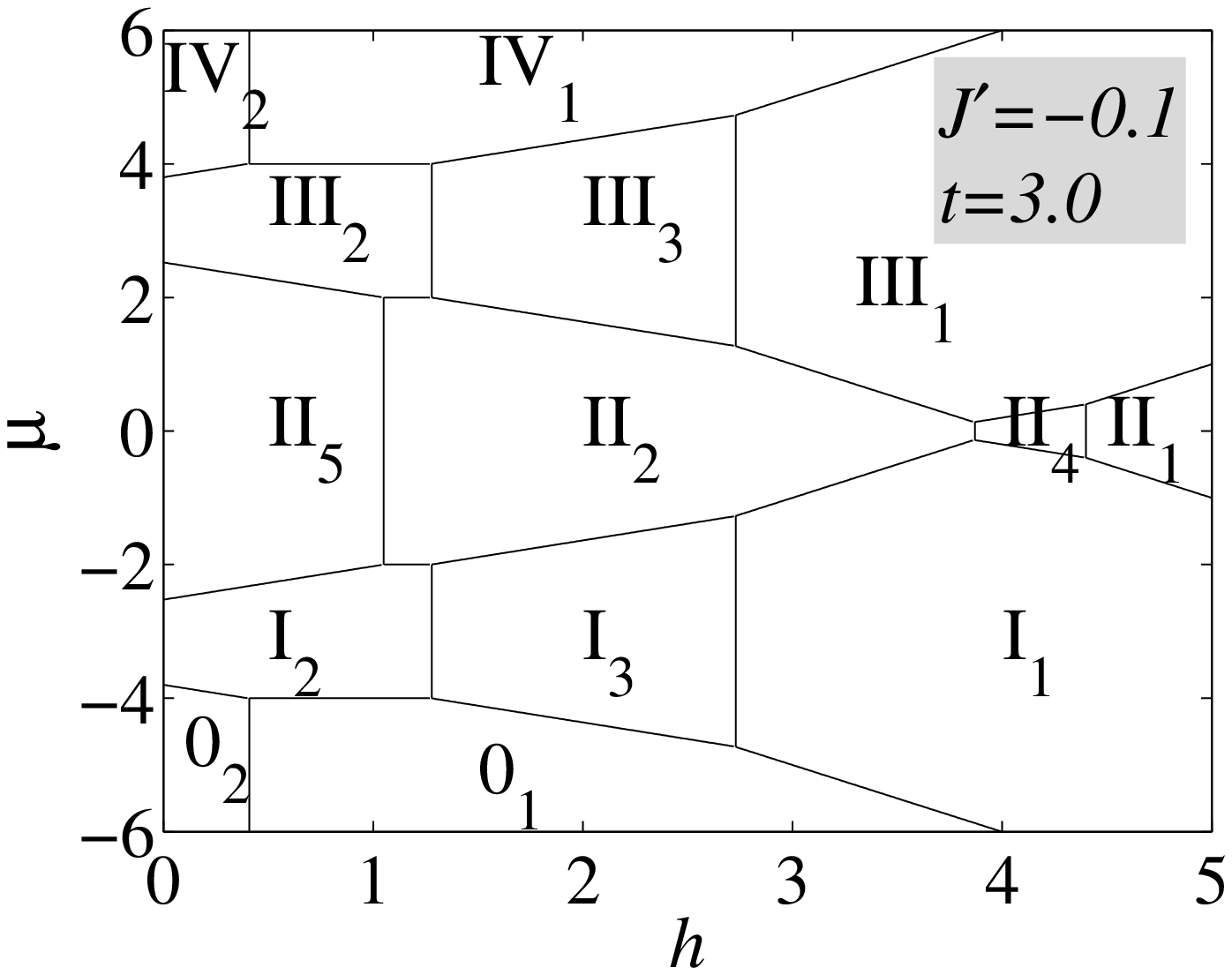}
\includegraphics[width=0.3\textwidth,trim=0 0 1.3cm 0.5cm, clip]{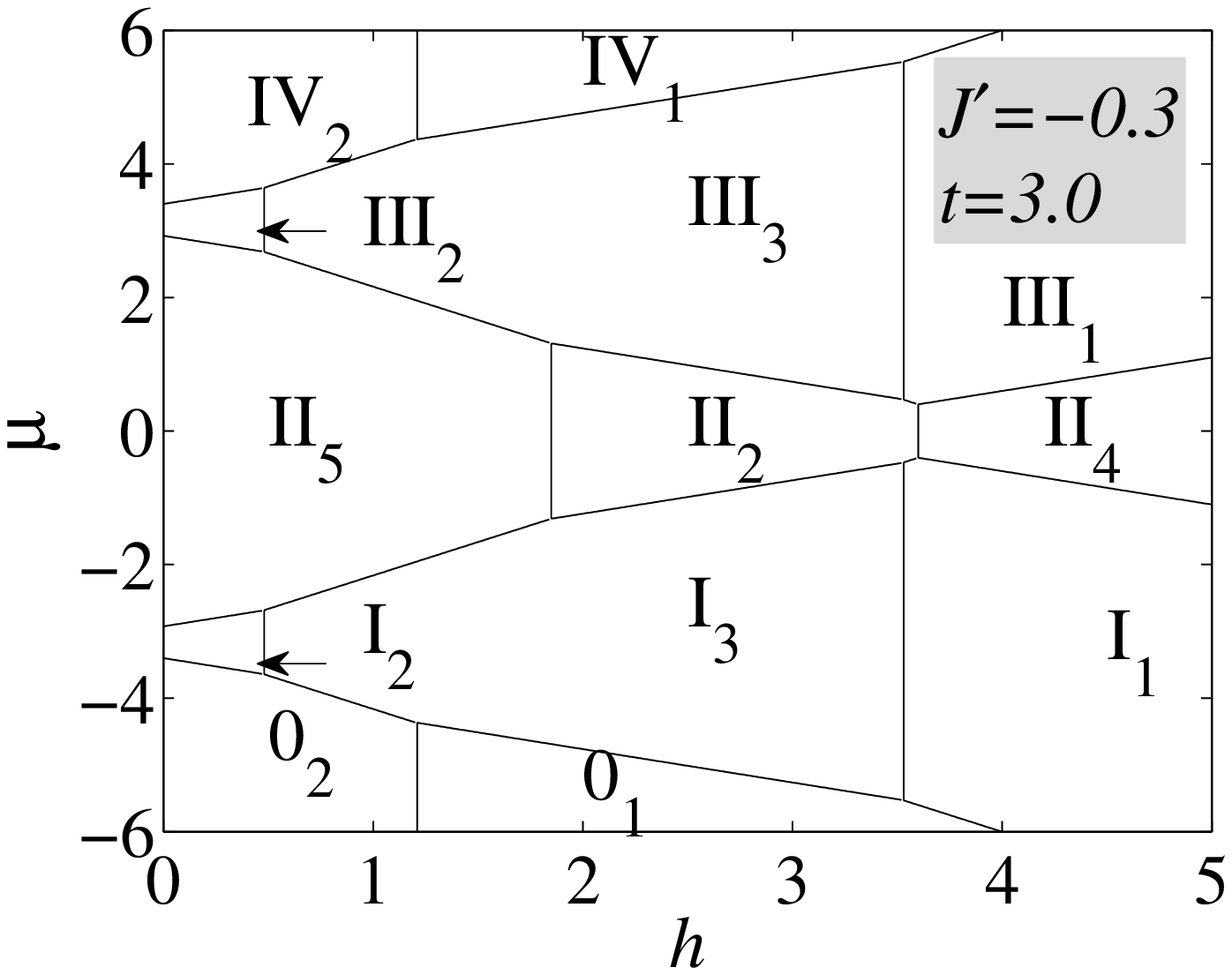}
\includegraphics[width=0.3\textwidth,trim=0 0 1.3cm 0.5cm, clip]{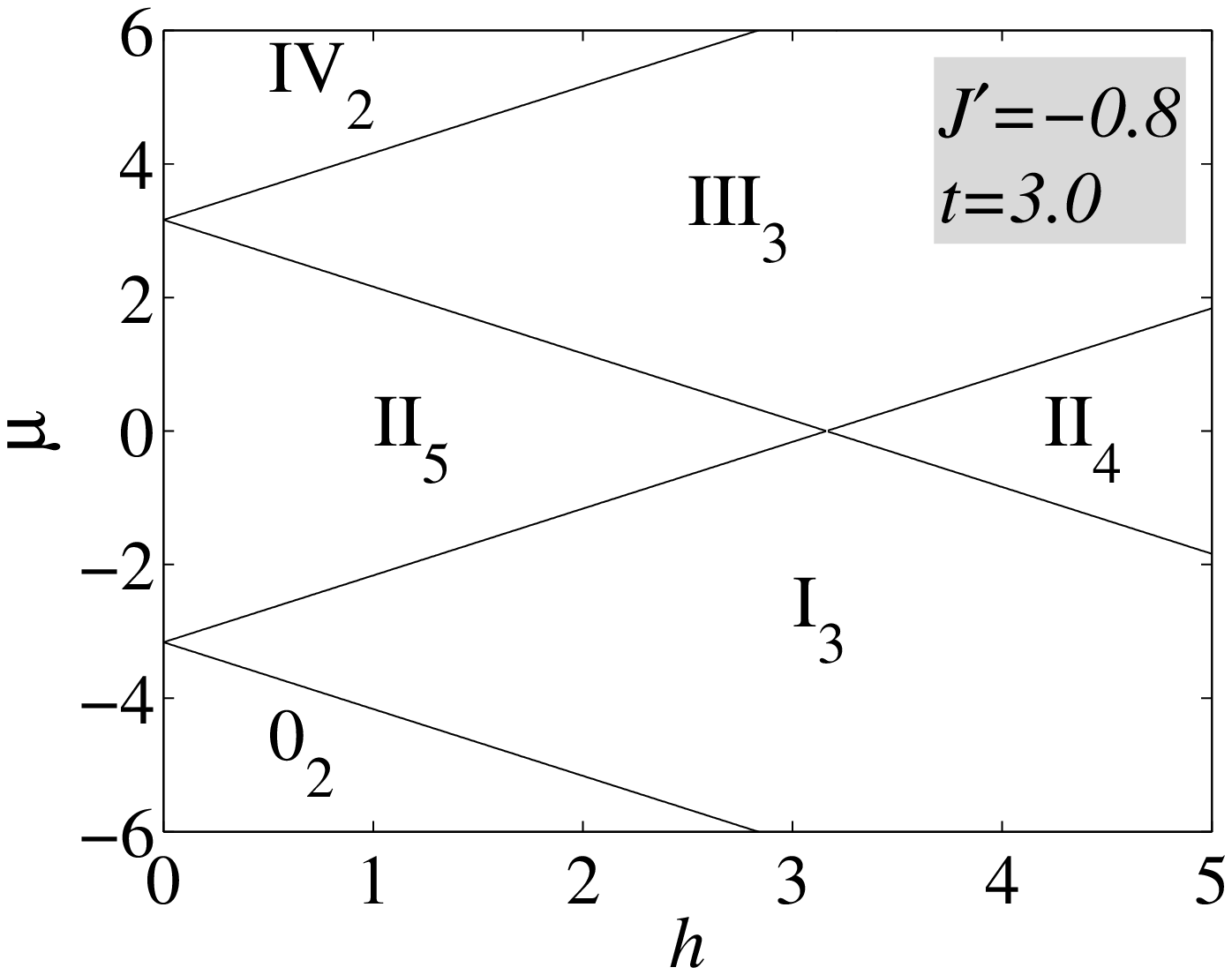}\\
\caption{\small Ground-state phase diagrams in the $\mu$-$h$ plane for  $J=1$ and selected values of $J'\leq0$ and $t$.}
\label{fig5}
\end{center}
\end{figure*}
The most interesting result of our investigations is the fact that the competing effect of the AF spin-spin coupling $J'<0$, the AF spin-electron coupling $J<0$, the hopping term $t$ and the magnetic field $h$ can produce various magnetic structures, which can be altered only by the changes of external magnetic field. 
It has been found that the AF spin-electron coupling $J<0$  leads to much higher diversity of magnetic structures in comparison with the F one ($J>0$)  and thus, it generates numerous field-driven phase transitions. Furthermore, the additional spin-spin interaction $J'$ may stabilize/produce selected magnetic structures, depending on  the character of the applied interaction, but the number of field-driven phase transitions is in general reduced. To complete our analysis, the remaining phase boundaries between the relevant phases have the following form:
\begin{eqnarray}
\hspace{-0.2cm}
\begin{array}{ll}
{\rm 0}_1{\rm -I}_2\;/\;{\rm III}_2{\rm -IV}_1:&\!\!\!\!\mu\!=\!(-1)^{u}(J-h+4h/q-t),\\
{\rm 0}_1{\rm -I}_3\;/\;{\rm III}_3{\rm -IV}_1:&\!\!\!\!\mu\!=\!(-1)^{u}(2(J'+h/q)-h-\sqrt{J^2+t^2}),\\
{\rm 0}_1{\rm -II}_3\;/\;{\rm II}_3{\rm -IV}_1:&\!\!\!\!\mu\!=\!(-1)^{u}(J-h+2h/q),\\
{\rm 0}_2{\rm -I}_3\;/\;{\rm III}_3{\rm -IV}_2:&\!\!\!\!\mu\!=\!(-1)^{u}(-h-\sqrt{J^2+t^2}), \\
{\rm I}_1{\rm -II}_3\;/\;{\rm II}_3{\rm -III}_1:&\!\!\!\!\mu\!=\!(-1)^{u}(3J-h+4h/q+t), \\
{\rm I}_1{\rm -II}_4\;/\;{\rm II}_4{\rm -III}_1:&\!\!\!\!\mu\!=\!(-1)^{u}(J+2J'-h+2h/q+t),\\
{\rm I}_2{\rm -II}_2\;/\;{\rm II}_2{\rm -III}_2:&\!\!\!\!\mu\!=\!(-1)^{u}(-J+h-4h/q-t),\\
{\rm I}_2{\rm -II}_3\;/\;{\rm II}_3{\rm -III}_2:&\!\!\!\!\mu\!=\!(-1)^{u}(J-h+t), \\
{\rm I}_2{\rm -II}_5\;/\;{\rm II}_5{\rm -III}_2:&\!\!\!\!\mu\!=\!(-1)^{u}(-J+h+t\\
&+2(J'-h/q-\sqrt{J^2+t^2})),\\
{\rm I}_3{\rm -II}_2\;/\;{\rm II}_2{\rm -III}_3:&\!\!\!\!\mu\!=\!(-1)^{u}(-2(J'+t+h/q)+h\\
&+\sqrt{J^2+t^2}), \\
{\rm I}_3{\rm -II}_3\;/\;{\rm II}_3{\rm -III}_3:&\!\!\!\!\mu\!=\!(-1)^{u}(2(J-J'+h/q)-h\\
&+\sqrt{J^2+t^2}),\\
{\rm I}_3{\rm -II}_4\;/\;{\rm II}_4{\rm -III}_3:&\!\!\!\!\mu\!=\!(-1)^{u}(-h+\sqrt{J^2+t^2}),\label{eqA5}\\
{\rm I}_3{\rm -II}_5\;/\;{\rm II}_5{\rm -III}_3:&\!\!\!\!\mu\!=\!(-1)^{u}(h-\sqrt{J^2+t^2}).
\end{array}
\hspace{-1cm}
\end{eqnarray}
Finally, the conditions for the last phase transitions complete our study:
\begin{eqnarray}
\begin{array}{ll}
{\rm II}_1{\rm -II}_4:&\!\!\!\! h/q\!=\!-J-J', \\
{\rm II}_2{\rm -II}_4:&\!\!\!\! h/q\!=\!(J'+t)/(q-1),\label{eqA6}  \\
{\rm II}_3{\rm -II}_4:&\!\!\!\! h/q\!=\!-J+J',\\
{\rm II}_4{\rm -II}_5:&\!\!\!\! h\!=\!\sqrt{J^2+t^2}.
\end{array}
\end{eqnarray}

\section{Conclusion}
\label{s4}
In conclusion, we have examined the influence of further-neighbor interaction on a diversity of magnetic structures in ground-state phase diagrams as well as  the number of field-driven phase transitions. It was found that the mutual interplay between the kinetic term, the Ising interaction between the localized spins and mobile electrons, the further-neighbor spin-spin interaction between the localized spins and the non-zero magnetic field leads to very rich magnetic phase diagrams including the F, AF as well as combined F-AF magnetic structures. Interestingly, it was found that the further-neighbor spin-spin interaction fundamentally influences the magnetic ground state and should be taken into account for a correct description of the magnetization processes of coupled spin-electron systems. In addition, it was observed that its inclusion strongly affects presence/existence of the field-induced phase transitions of metamagnetic nature at finite temperatures.  However, it is necessary to perform an extended theoretical analysis to answer this question satisfactorily. The work on this task is currently in progress~\cite{Cenci4}. 


\end{document}